\documentclass{IEEEtran}

\usepackage{xcolor}
\usepackage{graphicx}
\usepackage{ulem}
\usepackage{soul}
\usepackage{amsfonts}
\usepackage{amsmath}
\usepackage{amssymb}
\usepackage{tikz}
\usepackage{cite}

\title{Correcting spanning errors with a fractal code}

\author{Georgia M. Nixon and 
        Benjamin J. Brown
\IEEEcompsocitemizethanks{\IEEEcompsocthanksitem G. M. Nixon was with the Centre for Engineered Quantum Systems, School of Physics, University of Sydney, Sydney, New South Wales 2006, Australia. Her current affiliation is Cavendish Laboratory, University of Cambridge, J.J. Thomson Avenue, Cambridge, CB3 0HE, United Kingdom. \protect\\
E-mail: gmon2@cam.ac.uk
\IEEEcompsocthanksitem B. J. Brown is with the Centre for Engineered Quantum Systems, School of Physics, University of Sydney, Sydney, New South Wales 2006, Australia.}
}

\begin{document}

\maketitle

\begin{abstract}
The strongly correlated systems we use to realise quantum error-correcting codes may give rise to high-weight, problematic errors.
Encouragingly, we can expect local quantum error-correcting codes with no string-like logical operators --- such as the cubic code --- to be robust to highly correlated, one-dimensional errors that span their lattice.
 The challenge remains to design decoding algorithms that utilise the high distance of these codes.
 Here, we begin the development of such algorithms by proposing an efficient decoder for the `Fibonacci code'; a two-dimensional classical code that mimics the fractal nature of the cubic code.
 Our iterative decoder finds a correction through repeated use of minimum-weight perfect matching by exploiting symmetries of the code.
We perform numerical experiments that show our decoder is robust to one-dimensional, correlated errors.
First, using a bit-flip noise model at low error rates, we find that our decoder demonstrates a logical failure rate that scales super exponentially in the linear size of the lattice.
In contrast, a decoder that could not tolerate spanning errors would not achieve this rapid decay in failure rate with increasing system size.
We also find a finite threshold using a spanning noise model that introduces string-like errors that stretch along full rows and columns of the lattice.
These results provide direct evidence that our decoder is robust to one-dimensional, correlated errors that span the lattice.

\end{abstract}

\begin{IEEEkeywords}
classical coding theory, fractal codes, decoding algorithms, minimum-weight perfect matching, quantum error correction, fracton topological codes.
\end{IEEEkeywords}

\section{Introduction}

Quantum error-correcting codes~\cite{Dennis02, Terhal15, Brown16} enable us to reduce the error rate of encoded logical qubits arbitrarily. They are essential for the realisation of scalable quantum computation. Ideally, we will design codes that still function well when a large number of physical qubits experience errors. Such codes will be easier to construct as they will reduce the demand on the experimentalist to decrease the noise acting on the physical qubits of the system. We often consider using strongly correlated systems to construct a quantum computer~\cite{Nigg14, Kelly15, Takita16}. For such systems, it will be advantageous to search for codes that are robust to correlated errors~\cite{Fowler14, Nickerson17, Chubb18, Harper19}. Given such a code, it is important to find a decoding algorithm to interpret syndrome data to restore the system back to its encoded state with high probability. A good decoder will enable a code to reach its potential logical failure rate for a given system size.

The hardware we use to implement a quantum error-correcting code is often constrained by locality~\cite{Kelly15, Takita16}. It is well known that these constraints limit the scaling of important code parameters such as the code distance, $d$~\cite{Bravyi09, Bravyi10, Delfosse13}. The code distance is the support of the least-weight operator that can non-trivially transform an encoded state onto an orthogonal codeword. Broadly speaking, the code distance is proportional to the number of physical qubits of the code that can experience an error while the encoded information remains unaffected. For two-dimensional stabilizer codes of length $n =  \mathcal{O}(L \times L)$ such as the surface code~\cite{Dennis02, Kitaev03}, the distance is limited to $d = \mathcal{O}(L)$~\cite{Bravyi09, Bravyi10}. Moreover, the logical operators of these codes have string-like support and will not be robust to long string-like correlated errors.

We find a much richer space of quantum error-correcting codes by moving beyond two-dimensional systems. Notable examples include fracton topological codes~\cite{Chamon05, Castelnovo11, Haah11, Yoshida13a, Vijay16} and, in particular, type-II fracton codes~\cite{Castelnovo11, Haah11, Yoshida13a} such as the cubic code introduced by Haah~\cite{Haah11}. With the appropriate system size, these models encode $\mathcal{O}(L)$ logical qubits using $\mathcal{O}(L \times L \times L)$ physical qubits. This code rate scales equivalently to $L$ copies of the surface code. Unlike the surface code, type-II fracton codes have no string-like logical operators. This means that their distance scales super linearly in $L$. It also follows from this fact that these codes have the potential to be robust to correlated, one-dimensional errors. This is because the supports of their logical operators are much larger than a one-dimensional string. This is a qualitative advantage that type-II fracton topological codes have over the surface code. However, little work has been done to design decoders that can achieve a failure rate that scales exponentially in the code distance. 

Here, we propose a decoder for a classical code, called the Fibonacci code~\cite{Yoshida13a, Wolfram02}, that shares analogous properties with a type-II fracton code. Similar classical codes have also been shown to saturate known bounds on information storage capacity~\cite{Yoshida13}. In this sense, finding decoders for classical fractal codes is interesting in its own right. The Fibonacci code is a two-dimensional code with no string-like logical operators. Instead, the code has logical operators with fractal-like support. As such we look to develop decoders that can correct high-weight errors. We find that we can use minimum-weight perfect matching~\cite{Edmonds65, Kolmogorov09} to pair defects on fractal subsets of the lattice to test whether specific physical bits have experienced errors. We use this observation to propose an iterative decoder that enables us to correct errors with string-like support. To demonstrate this feature we provide numerical results indicating that the logical failure rate of our decoder decays super exponentially in $L$. This suggests that there are no errors of weight $\sim L$ that will cause our decoder to fail. In addition to this, we test our decoder using a correlated error model that introduces one-dimensional errors that span the system along horizontal and vertical lines of the two-dimensional lattice. The positive threshold that we demonstrate provides further evidence that our decoder is robust to correlated errors that span the lattice.

The decoder we present has a parallelised runtime complexity of $\mathcal{O}(L^{3D})$ where $D = 1+\log_2 \phi \approx 1.69$ and $\phi $ is the golden mean~\cite{Yoshida13a, Devakul19}. Specifically, we make $\mathcal{O}(L^2)$ calls to the minimum-weight perfect matching algorithm, the complexity of which is $\mathcal{O}(V^3)$~\cite{Gabow73, Lawler76} where the number of vertices is $V = \mathcal{O}(L^D)$ in our model. Calls to the minimum-weight perfect matching algorithm can be parallelised to reduce the runtime complexity of our decoder. We note that the minimum-weight perfect matching algorithm can be replaced with other paring subroutines, for instance~\cite{Delfosse17}, which has a runtime of almost $\mathcal{O}(V)$ with, presumably, a small cost in performance.

The remainder of the Manuscript is organised as follows. In Sec.~\ref{Sec:Code} we define the Fibonacci code and describe its novel features, then in Sec.~\ref{Sec:Decoder} we introduce the decoder we will use. In Sec.~\ref{Sec:Results} we present our numerical results. After offering some concluding remarks, in Appendix~\ref{App:Runtime} we discuss the runtime of the decoder, in Appendix~\ref{App:Methods} we discuss the analysis of our numerical data, and in Appendix~\ref{App:OtherDecoders} we consider adaptations of our decoding methods to other classical codes and compare their properties with those of the Fibonacci code.

\section{The Fibonacci code}
\label{Sec:Code}

We define the Fibonacci code on a two-dimensional lattice such that it encodes $k = L$ logical bits using $n = L^2 / 2$ physical bits. The exact distance of the code, $d > L$, is unknown. We begin by introducing some basic notation that we will use throughout this work. We will describe the basic features of the Fibonacci code~\cite{Yoshida13a, Wolfram02}, its logical operators and how it responds to errors. We will also review the symmetries of the model that will be important in Sec.~\ref{Sec:Decoder} where we introduce the decoder.

\subsection{The lattice, stabilizers and logical operators}

\begin{figure}[b]
\begin{center}
	\includegraphics{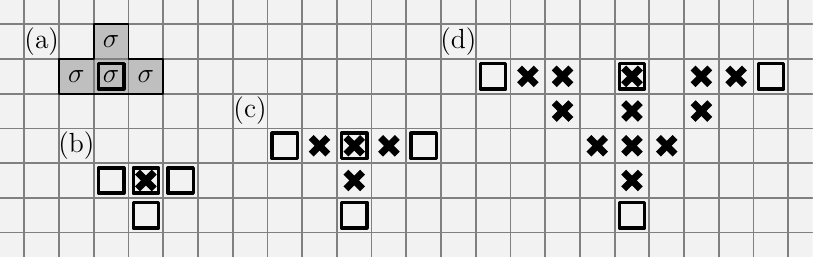}
	\end{center}
	\caption{\label{Fig:Stabilizer} The Fibonacci code. (a)~A parity check $S_f$ is a weight four term where the face $f$ is marked by a black box. Errors at~(b), (c), and (d)~are marked by black crosses and cause the same syndrome at different length scales. The violated parity checks are marked by black boxes.}
\end{figure}

The Fibonacci code is defined with a single bit $\sigma_f = \{ 0,\, 1 \}$ on each of the faces $f$ of a square lattice with periodic boundary conditions. We focus on lattices with dimensions $L \times L/2$ where $L = 2 ^ N$ for integers $N \ge 2$, but we give a short discussion on how we might generalise our results beyond this case in SubSec.~\ref{SubSec:Repeats}. We specify faces using coordinates $f = (x,y)$ with integers $ 1 \le  x \le L$ and $1 \le y \le L/2$ where $f + L\hat{x} = f + L \hat{y}/2 = f$ for all $f$ due to the periodic boundary conditions and where $\hat{x} = (1,0)$ and $\hat{y} = (0,1)$ are canonical unit vectors. It will be helpful to define the functions $X(r) = x$ and $Y(r) = y$ for coordinates $r = (x,\,y)$.

The code is defined by its parity checks, or `stabilizers'. The stabilizers are defined as follows:
\begin{equation}
S_f = \sigma_{f+\hat{y}} \oplus \sigma_{f-\hat{x}} \oplus \sigma_{f} \oplus \sigma_{f+\hat{x}}.
\label{Eqn:Stabilizer} 
\end{equation}
where `$\oplus$' denotes addition modulo 2. All codewords satisfy $S_f = 0$ for all $f$. We show a stabilizer in Fig.~\ref{Fig:Stabilizer}(a); its central face $f$ is shown with a black box.

\begin{figure}
\begin{center}
	\includegraphics{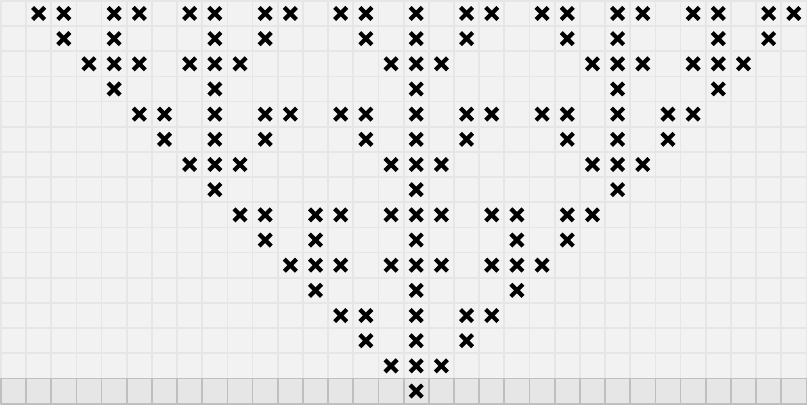}
	\end{center}
	\caption{\label{Fig:Logical} A codeword of the Fibonacci code. We show a lattice of size $L \times L/2$ physical bits with $L = 32$ where bits flipped to the $1$ state are marked by black crosses, and otherwise take the value $0$. We can encode $L$ bits using the Fibonacci code. Codewords are produced by first writing an arbitrary bit string of length $L$ onto the shaded bottom row. The remaining rows are then determined with the cellular automaton update rules of Eqn.~(\ref{Eqn:Update}). The locations of bits in the 1 state in this codeword reveal the fractal structure.}
\end{figure}

The $2^k$ codewords of the Fibonacci code are found using cellular automaton update rules~\cite{Wolfram02} over the rows of the lattice, see Fig.~\ref{Fig:Logical}. We encode an arbitrary bit string of length $k = L$ on the bottom row of the lattice which includes all $\sigma_f$ where $Y(f) = 1$. This row is shaded in Fig.~\ref{Fig:Logical}. We then determine the values $\sigma_f$ on the second row where $Y(f) = 2$ using the update rule 
\begin{equation}
\sigma_{f} = \sigma_{f-\hat{x}-\hat{y}} \oplus \sigma_{f-\hat{y}} \oplus \sigma_{f+\hat{x}-\hat{y}},  \label{Eqn:Update}
\end{equation}
before proceeding onto the third row and so on.
This update rule is generated using the stabilizer definition for the Fibonacci code given in  Eqn.~(\ref{Eqn:Stabilizer}). If $S_f = 0$, we have that $ \sigma_{f+\hat{y}} \oplus \sigma_{f-\hat{x}}  \oplus \sigma_{f} \oplus \sigma_{f+\hat{x}}  = 0$. Adding $\sigma_{f+\hat{y}}$ to both sides of this relation and using $\sigma_f \oplus \sigma_f = 0$ for all $f$ we recover Eqn.~(\ref{Eqn:Update}).
Assuming the correct system sizes with $L = 2^N$, codewords obtained with these update rules will satisfy the constraints of the parity checks once all the rows have been updated.

\subsection{Errors}

The fractal structure of the Fibonacci code is reflected by its response to errors. We begin with a valid codeword of the Fibonacci code that satisfies all of the stabilizer constraints. If bit-flip errors $\sigma_f \rightarrow \sigma_f \oplus 1$ occur, some of the stabilizer constraints may be violated. For a given error configuration, we say that there is a defect at face $f$ if $S_f = 1$. The list of all the stabilizers $S_f$ for a given error configuration is called the error syndrome.

Let us now look at the fractal structure of the Fibonacci code. To do so, we consider error configurations that produce a syndrome with a self-similar pattern as we move single defects to isolated locations on the lattice, see Fig.~\ref{Fig:Stabilizer}(b-d). In Fig.~\ref{Fig:Stabilizer}(b) we first show the syndrome of four defects generated by an error acting on a single bit. Then, in Fig.~\ref{Fig:Stabilizer}(c) we show an error configuration resulting in four defects where now three defects have been displaced a single lattice site away from $f$, while a single defect remains at $f$. One can verify that we cannot delocalise a single defect without separating the other defects over the lattice by a proportionate distance~\cite{Haah11}. In this sense, the syndrome respects a self-similar pattern. In Fig.~\ref{Fig:Stabilizer}(d) we show the three defects separated by three lattice spacings from the central face $f$. If we continue to separate the defects over the entire lattice by introducing more errors, they will eventually align on the same face and introduce an undetectable logical error, as shown in Fig.~\ref{Fig:Logical}. The support of the flipped bits on the lattice shows the fractal-like nature of the codewords of the Fibonacci code.

As an aside, we remark that fractal codes were initially of interest due to the energy landscape of the corresponding Hamiltonian $H = \sum_f S_f$, where the energy is proportional to the number of defects on the lattice. This Hamiltonian gives rise to glassy dynamics~\cite{Chamon05, Castelnovo11, Newman99} and, as such, has been of interest for the study of self correction~\cite{Brown16, Castelnovo11, Haah11, Bravyi13b}. As these codes satisfy the no-strings rule~\cite{Haah11} it is known that a local noise model must overcome an energy barrier that is logarithmic in the system size to introduce a logical error~\cite{Bravyi11a}. For instance, if we were to look at a sequence of configurations $\sigma_f(t)$ with $t = 0,\,1,\,2,\dots,\, T$ such that $\sigma(t)$ differs from $\sigma_f(t-1)$ by a small constant number of bit flips, i.e., the hamming weight of $\sigma(t) \oplus \sigma(t-1)$ is small for all $t$, and $\sigma(T)$ is a codeword with $\sigma(T) \not= \sigma(0)$, then the energy of $H$ must become very large at some point along this sequence due to the energy barrier of the Hamiltonian. See Ref.~\cite{Brown16} and references therein for a discussion on energy barriers and their physical significance. Intuitively, this is due to the fractal support of the faces that must experience bit flips to delocalise the defects over a long distance as shown in Figs.~\ref{Fig:Stabilizer}(b-d). Further work has shown that the dynamics of a fractal code in a thermal environment are partially self correcting~\cite{Bravyi13b} due to their logarithmic energy barrier.

\subsection{Fundamental symmetries}

\begin{figure}
\begin{center}
	\includegraphics{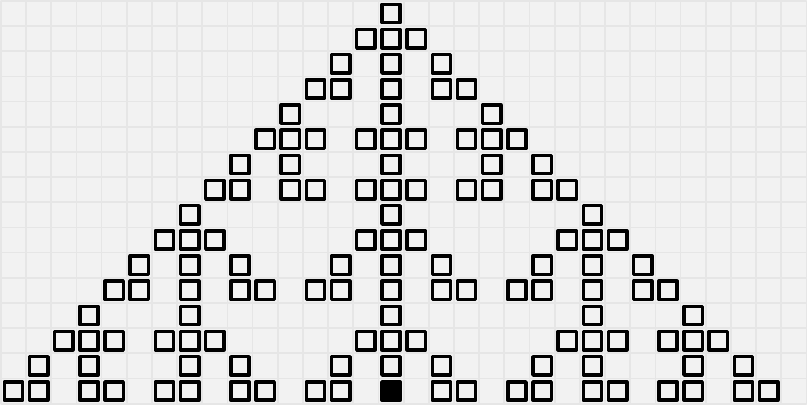}
	\end{center}
	\caption{\label{Fig:Symmetry} The symmetry $\Sigma_F$ for system size $L = 32$.  Stabilizers $S_f \in \Sigma_F$ are marked with black boxes. The box on face $F$ is filled black. The sum of all parity checks $S_f \in \Sigma_F$ returns $0$. This implies that, for any error, there will always be an even number of defects on the faces of the symmetry. We will use this symmetry, and spatial translations thereof, to design our decoder.}
\end{figure}

We finally define the symmetries of the Fibonacci code~\cite{Devakul19}. A symmetry $\Sigma$ is a subset of the stabilizers $S_f$ that satisfies
\begin{equation}
\bigoplus_{f \in \Sigma} S_f = 0. \label{Eqn:Symmetry}
\end{equation}
Importantly, the definition of a symmetry implies that the syndrome configuration caused by any arbitrary error will introduce an even number of defects on the subset of faces $S_f \in \Sigma$. This can be regarded as a conservation law among the defects of the code. We will exploit this feature of the symmetry in the next section where we propose a decoder.

Here we define a special subset of symmetries that we refer to as `fundamental' symmetries for our decoding algorithm. An example of a fundamental symmetry is shown in Fig.~\ref{Fig:Symmetry}. The faces $f$ identifying the stabilizers $S_f$ included in this symmetry are marked by black squares. We label this fundamental symmetry $\Sigma_F$ where $F = (L/2, 1)$ is marked by the filled black square in the figure. Also note that $S_F \in \Sigma_F$. Other fundamental symmetries are the $L^2/2$ possible spatial translations of that shown in Fig.~\ref{Fig:Symmetry}.
	
We find $\Sigma_F$ using cellular automaton update rules, see Refs.~\cite{Yoshida13, Devakul19}. These rules specify a bit $b_f$ for each face of the lattice defined such that $b_f = 1$ if the stabilizer $S_f$ is included in the symmetry and $b_f = 0$ otherwise. Let us first specify the bit string for the top row of the lattice, i.e., bits $b_f$ with $Y(f) = L/2$. We impose that $b_f = 0$ for all top row bits apart from a single bit at face  $f = (L/2, L/2) $  with $ b_{(L/2,L/2)}= 1$. We then determine all bit values $b_f$ on the second row where $Y(f) = L/2 - 1$ using the update rule
\begin{equation}
b_f = b_{f  - \hat{x}+\hat{y}} \oplus  b_{f +\hat{y}}  \oplus  b_{f + \hat{x} + \hat{y}},
\label{Eqn:CellularAutomata}
\end{equation} 
where addition is taken modulo two. Using Eqn.~(\ref{Eqn:CellularAutomata}), we  evaluate all bits on each row sequentially in descending order until all rows of the lattice have been updated. Once we have updated all of the rows in this order, we recover a fundamental symmetry for lattices of size $L \times L/2$ where $L = 2^N$ and $N \ge 2$ is an integer. Given $\Sigma_F$, we then find $\Sigma_{F+r}$ with $r = (x, y)$ some translation such that $S_{f'} \in \Sigma_{F+r}$ if and only if $S_{f'} = S_{f+r}$ where $S_{f} \in \Sigma_F$.

For completeness, we finally remark that the symmetries of a code form an Abelian group $ \Sigma \in \mathcal{S}$. Let us define this group in terms of the bit strings $b_f$ we defined above that specify group elements $\Sigma$. The bit string $b_f = 0$ for all $f$ specifies the identity element of the group, $ \Sigma_1 = \emptyset$. The binary operation of the group $ \Sigma'' = \Sigma \circ \Sigma'$ is the bitwise addition of the bit strings specifying each of the symmetries, namely, $\Sigma''$ is specified by the bit string $b_f'' = b_f \oplus b_f'$ where $b_f $ ($b_f'$) specify the elements of $\Sigma$ ($\Sigma'$). All elements of the group are self inverse. One can check that the group of symmetries can be generated by $L$ elements, for instance, the fundamental symmetries $\Sigma_{F+j \hat{x}}$ where $ 1 \le j \le L$. This is consistent with the dimensionality of the code space of the Fibonacci code we study here.

\section{Decoder}
\label{Sec:Decoder}

\begin{figure}
\begin{center}
	\includegraphics{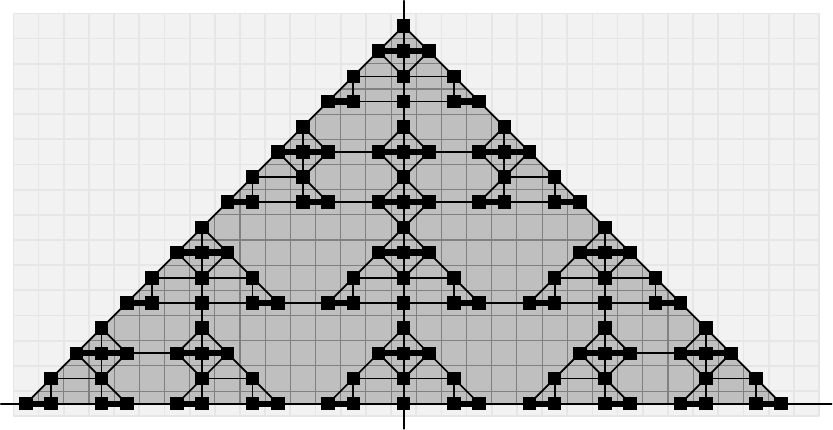}
	\end{center}
	\caption{\label{Fig:MatchingGraph} The matching graph. The faces that support stabilizers $S_f \in \Sigma_F$ are marked by filled black boxes. Pairs of faces that each support a defect following the introduction of a single bit-flip error are connected by edges. Thick edges connect pairs of faces where there are two single-bit errors that can create the given pair of defects. With only two exceptions, all of the edges are supported on the dark grey triangle. The other two edges connect two opposite points of the triangle, horizontally and vertically, around the periodic lattice.}
\end{figure}

Here we describe our decoder for the Fibonacci code. With the appropriate choice of symmetries, we find two distinct tests to determine if a physical bit of the lattice has experienced an error. Our decoder dynamically updates the correction according to the outcomes of the tests that we use on each of the bits of the code. In what follows, after explaining how we use the symmetries of the system together with the minimum-weight perfect matching algorithm, we go on to describe how we use these components to design two tests that we call the horizontal probe and the vertical probe to determine if a bit has experienced an error. We then discuss why we expect that the probes can deal with high-weight errors before we propose an iterative decoder using the probes we have introduced.

\subsection{The matching graph}

\label{Subsec:MatchingGraph}

In Ref.~\cite{Brown19}, it is proposed that we can design a decoder given sufficient knowledge of the symmetries of the code. Specifically, the defects an error produces must respect a conservation law for each of its symmetries. For local codes, this defect conservation can be likened to Gauss' law~\cite{Kitaev03}. Taking a more practical perspective, for some symmetry, we find that errors produce an even number of defects whose separation grows with the weight of the error. With this observation, it is natural to propose a decoder by pairing nearby defects on the stabilizers of a symmetry to determine the most likely errors that caused a given syndrome. We now examine the symmetries of the Fibonacci code to determine a strategy for decoding.

We define a matching graph $\mathcal{G}_F$. It will be used for the pairing procedure carried out by the minimum-weight perfect matching subroutines. In Fig.~\ref{Fig:MatchingGraph}, we mark the faces that support a stabilizer of a fundamental symmetry of the Fibonacci code $\Sigma_F$ with black squares. We assign each of these faces with a vertex of the matching graph. Pairs of vertices of the symmetry are connected with an edge if a single error can produce a defect on both of the vertices of the pair. 
The errors that give rise to each edge are shown in Fig.~\ref{Fig:EdgeErrors}. We will also require translations of the matching graph we have just defined. The matching graph $\mathcal{G}_{F+r} $ corresponds to the set of vertices specified by the faces of the fundamental symmetry $\Sigma_{F+r}$.

\subsection{Minimum-weight perfect matching}

\begin{figure}
\begin{center}
	\includegraphics{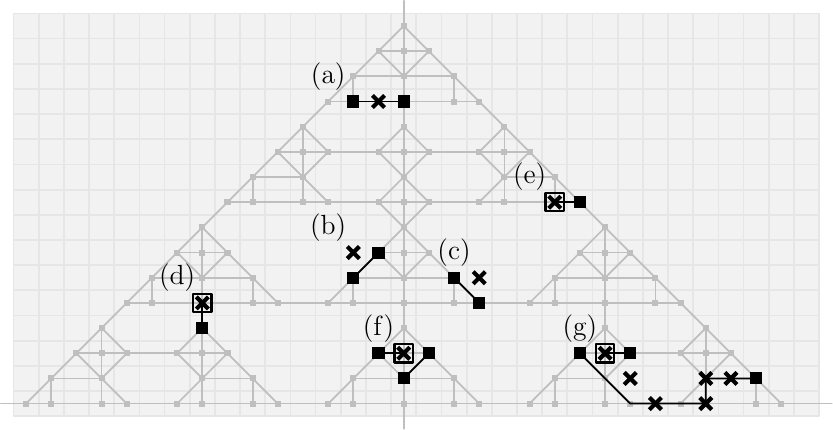}
	\end{center}
	\caption{\label{Fig:EdgeErrors} Errors, marked by black crosses, can be viewed as string segments that correspond to the edges of Fig.~\ref{Fig:MatchingGraph}. (a)~A single error that introduces a string segment connecting two faces separated by one lattice spacing. In~(b)~and (c)~we show errors that give rise to diagonal strings. (d)~An error that corresponds to a vertical string. (e)~A degenerate error; an error on the marked face, or the face to its right, will both introduce this pair of defects and the corresponding error string. We account for this degeneracy in the minimum-weight perfect matching subroutines. (f)~An error that introduces four defects. We can regard this error as a pair of short strings with defects at the endpoints of both strings. The results of our work show this does not cause a significant issue for the minimum-weight perfect matching algorithm. (g)~In general, many errors can combine to give longer strings on the graph where defects lie at the end points of the string.}
\end{figure}

Errors can be likened to small string segments that lie on the edges of the matching graph, where defects lie at their endpoints, see Figs.~\ref{Fig:EdgeErrors}(a-f). In general, multiple errors can combine to create longer string segments, see Fig.~\ref{Fig:EdgeErrors}(g). By assuming a noise model that introduces low-weight errors, we expect that an error configuration will give rise to short strings with defects at their endpoints. With this observation, our decoder pairs nearby defects on a given symmetry, that are local with respect to the edges of the matching graph, to estimate the error that has occurred. 

We use the Kolmogorov implementation~\cite{Kolmogorov09} of the minimum-weight perfect matching algorithm due to Edmonds~\cite{Edmonds65} to pair defects supported on a symmetry. The algorithm takes a graph of nodes with weighted connections and returns a `matching', i.e. a set of node pairs from the original graph such that the sum of connection weights between pairs is minimised. This implementation has worst case complexity of $\mathcal{O}(V^3E)$ where $V$ is the number of input nodes and $E = \mathcal{O}(V^2)$ is the number of input connections for a complete graph.

At this point, we note that we have used unconventional terminology to describe the input graph and output matching of the minimum-weight perfect matching algorithm. Specifically, we have used the synonyms `connections' and `nodes' in lieu of `edges' and `vertices'. This is to avoid confusion with the edges and vertices of the matching graph defined in SubSec.~\ref{Subsec:MatchingGraph}. We will maintain this convention throughout our exposition. We will say that there is a `matching' between two nodes if they share a connection in the output of the minimum-weight perfect matching algorithm.

We must create an input graph for the minimum-weight perfect matching subroutine such that the connections of the output matching align closely to the error strings. We make each defect supported on the fundamental symmetry a node of the input graph. We then assign weights to the connections between pairs of nodes according to, approximately, the logarithm of the probability that a local collection of error events created that pair of defects. Assuming the independent and identically distributed noise model, to first order, this is proportional to the shortest path between any two given defects with respect to the matching graph.

We improve our decoding algorithm by assigning weights to connections that account for degeneracy. A degenerate syndrome is one where more than one error configuration may have given rise to the same pattern of defects. See, for instance, Fig.~\ref{Fig:EdgeErrors}(e) where the same pair of defects shown on the fundamental symmetry can be produced by a single bit flip at two locations. When  considering subsets of the syndrome that intersect with the fundamental symmetries of the code, we can account for degeneracy between defects with an appropriate choice of connection weight function. Given two defects at locations $f_1$ and $f_2$ of the matching graph, we weight their connection according to the formula
\begin{equation}
W(f_1,f_2) \propto  - \log D + E \log \left( \frac{1-p}{p} \right) \label{Eqn:EdgeWeights}
\end{equation}
where $E$ is the length of the shortest path separating $f_1$ and $f_2$ along the matching graph and $D$ is the number of least-weight errors that can create the pair of defects. For references on how to account for degeneracy using minimum-weight matching, see~\cite{Stace10, Criger18}.

\subsection{Error detection probes}

\begin{figure}
\begin{center}
	\includegraphics{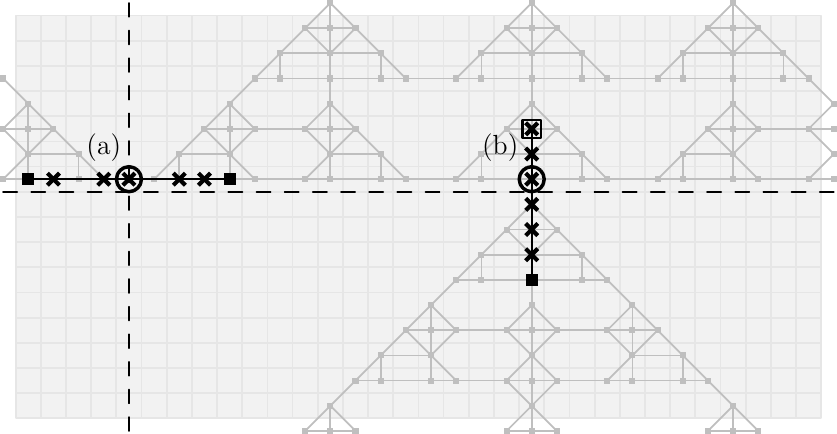}
	\end{center}
	\caption{\label{Fig:Probes} A matching graph of the Fibonacci code. Supposing low-weight errors, if the matching pairs two defects with a connection that crosses the dashed vertical line, then there must be an error on the horizontal edge that intersects the dashed vertical line at the circled special site (a). Likewise, there is a unique face circled at (b) that must support an error in order to produce a string segment that crosses the dashed horizontal line.}
\end{figure}

We can probe the physical bits of the Fibonacci code for errors individually by pairing the defects lying on the faces of the fundamental symmetries. We exploit the following observation to find a correction for the code. With few exceptions, all short paths that pair defects on the matching graph are supported entirely on the dark grey triangle shown in Fig.~\ref{Fig:MatchingGraph}. This is because all but two edges of the matching graph are supported on this region. We disregard defects that are paired via paths supported on edges that lie on the grey triangle.

There are exactly two edges of the matching graph that connect opposite sides of the shaded triangle in Fig.~\ref{Fig:MatchingGraph}. These are special edges of the matching graph. We show these edges on a translated matching graph $\mathcal{G}_f$ in Fig.~\ref{Fig:Probes}. One edge intersects the vertical dashed line defined on coordinates $c = (x,y)$ such that $X(c) = X( f + \hat{x}L/2 ) $ where we recall that $f$ indexes the central face of the matching graph, see Fig.~\ref{Fig:Probes}(a). The other edge intersects the horizontal dashed line, see Fig.~\ref{Fig:Probes}(b). The equation for this line is aligned with the coordinates such that $Y(c) = Y(f - \hat{y}/2)$.

We now look at the special edges that intersect the dashed lines in Fig.~\ref{Fig:Probes} more closely. In both cases, in order to create a pair of defects with a low-weight error that is connected by a short path on the matching graph that includes a special edge, then we find that one specific bit must have experienced an error. We mark these bits with circles in Figs.~\ref{Fig:Probes}(a) and~(b). To make this observation more explicit, we show short error strings in Figs.~\ref{Fig:Probes}(a) and~(b) which have errors on the faces marked by a circle. Crucially, if we assume low-weight errors, the pairing the minimum-weight perfect matching subroutine returns will connect defects via a short path that crosses the dashed lines if and only if the respective bits at the special faces, marked by circles, have experienced errors. It follows that if the minimum-weight perfect matching algorithm returns a matching that crosses these dashed lines, then it is highly-likely that an error has occurred at these special sites.

We can use the observations discussed above to test each of the bits of the code for errors individually. In fact, we have two separate tests for each bit. We call these the horizontal probe $H(f) = 0,\,1$ and the vertical probe $V(f) = 0,\,1$ where the probes return $0$ if they do not suspect that a face $f$ has experienced an error and $1$ otherwise. Let us define these tests explicitly. The horizontal probe performs minimum-weight perfect matching on the defects lying on the matching graph $\mathcal{G}_f$ with vertices corresponding to the faces defined by the fundamental symmetry $\Sigma_f$. The probe $H(f)$ returns the parity of the number of edges returned by the minimum-weight perfect matching algorithm that cross the horizontal dashed line with coordinates $c$ satisfying $Y(c) = Y(f - \hat{y}/2)$. Likewise, we can test face $f$ for a bit flip with the vertical probe by using minimum-weight perfect matching to pair defects on the vertices of the matching graph $\mathcal{G}_{f+ \hat{x} L/2}$. The vertical probe predicts a bit flip at $\sigma_f$, i.e. returns $V(f) = 1$, if and only if an odd parity of matchings cross the vertical dashed line $X(c) = X(f)$.

\begin{figure}[b]
	\begin{center}
		\includegraphics{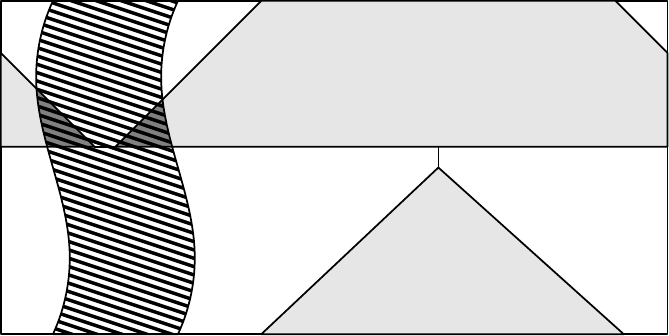}
	\end{center}
	\caption{\label{Fig:Intuition} A high-weight error supported on the hatched region that covers a vertical span of the lattice. The light grey triangle shows a symmetry that has been translated such that the vertical probe tests a face $f$ lying on the hatched region for an error. Although there are many errors on the hatched region, it will be straightforward for the vertical probe to determine faces that support errors. If an error lies on the face $f$ that is being tested, then there must be an odd number of defects on the two separate dark grey regions where the symmetry and the hatched region overlap. Such a configuration will mean an edge will connect the two disjoint dark grey regions, indicating an error at $f$ using the vertical probe.}
\end{figure}

\subsection{Correcting high-weight errors}

We have established two different tests to determine if $\sigma_f$ has been flipped. We expect that at least one of the two probes to identify the locations of errors accurately, even if an error configuration spans the lattice. We illustrate the mechanism that enables this in Fig.~\ref{Fig:Intuition}. Here we show a hatched region that spans the lattice vertically where we suppose that many errors have occurred. If there are many errors on this band, the horizontal probe will inaccurately estimate the locations of errors on this region. However, the vertical probe will successfully identify the locations of errors with high likelihood.

The grey triangle in Fig.~\ref{Fig:Intuition} depicts the matching graph that has been aligned to use the vertical probe $V(f)$ to test some bit $\sigma_f$ on the hatched region for an error. While there are many errors in this region, the overlap of the error with the symmetry is relatively small. We shade the overlap of the symmetry with the error in dark grey. The shaded area is separated into two parts. If there is no error on the face that is being tested, then there must be an even number of defects in each of the two parts of the shaded region. Otherwise, there will be an odd number of defects lying on the neighbourhood of each of these regions such that the minimum-weight perfect matching subroutine will pair two defects from the two dark grey regions. The connection of the pair will cross the line where $c = X(f)$, indicating that an error has occurred at $f$.

We have proposed two probes $H(f)$ and $V(f)$ to determine if $\sigma_f$ has been flipped. The success of these probes will depend on the incident error configuration. In general, it is not clear which of the two probes will perform best. In the example we have considered, the vertical probe will identify errors on the hatched region with high probability. However, it is easy to imagine error configurations where the horizontal probe will be better suited to identify errors. A more reliable test might require that both $H(f)$ and $V(f)$ agree an error has occurred at $f$, but in some cases this might not be the best strategy. Consider again the example shown in Fig.~\ref{Fig:Intuition}. In this example $V(f)$ will determine the errors very reliably, but given that $H(f)$ will be unreliable at correcting this error it may be too stringent to correct a bit only if both $V(f)$ and $H(f)$ register an error at $f$. To this end in the following section we propose a decoder that finds a correction iteratively where the results of different tests are chosen dynamically at different stages of decoding.

\subsection{Finding a correction using the probes}

\begin{figure}
	\begin{center}
		\includegraphics{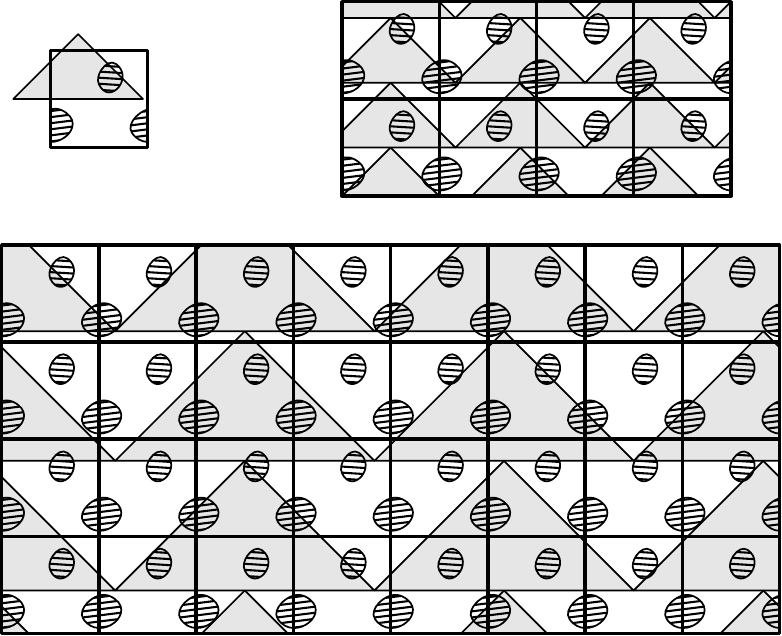}
	\end{center}
	\caption{\label{Fig:Repeats} Consider a lattice with irregular dimensions such that a fundamental symmetry cannot be found~(top left). We can duplicate the syndrome such that regular symmetries can be found whereby probes can be identified to find the errors of the duplicated system (top right). It may be beneficial to repeat the syndrome beyond the smallest duplicated system that is compatible with our proposed method. This is because a larger system may be better suited to identify the locations where errors have occurred (bottom).}
\end{figure}

Given a syndrome with stabilizers $S_f =\{0,\, 1\}$ we look for a correction $C_f$ that will correct all the bits that have experienced errors. The correction is a bit string $C_f = \{0,\, 1 \}$ for all faces $f$ of the lattice. The correction will flip $\sigma_f$ if and only if $C_f = 1$. To determine $C_f$ we choose between three different proposed corrections. The first, $C_f^V$, is evaluated with the vertical probe where $C^V_f = V(f)$. The second, $C_f^H$, is evaluated with the horizontal probe, i.e., $C^H_f = H(f)$. The final correction, $C_f^D$, is evaluated using both probes, $C_f^D = V(f)  H(f) $.

Once the three corrections $C_f^V$, $C_f^H$ and $C_f^D$ have been determined, we must choose which of the three corrections to apply. We assume that a good strategy to find the best correction will be to minimise the weight of the new syndrome once the correction has been applied. We evaluate the syndrome for each of the correction operators, i.e., 
\begin{equation}
S^\Gamma_f = C^\Gamma_{f+\hat{y}} \oplus C^\Gamma_{f-\hat{x}} \oplus C^\Gamma_{f} \oplus C^\Gamma_{f+\hat{x}}, \label{Eqn:Stabilizer2}
\end{equation}
where $\Gamma = V, H ,D$. We then choose the correction that minimises the weight of the syndrome after the correction has been applied. Specifically, we take $C_f = C^\Gamma_f$ for $\Gamma = V,\,H $ or $D$ for the updated syndrome $\tilde{S}_f^\Gamma = S_f \oplus S_f^\Gamma$ where the bit string  $\tilde{S}^\Gamma$ has the smallest number of checks with $\tilde{S}_f^\Gamma = 1$. Once we have determined a new correction, we update the error configuration with the correction we have obtained, and update its syndrome.

In general, we do not expect to correct all errors with one iteration of the decoder. If parity checks remain violated after one decoder sequence, we continue with another iteration of the decoder. If after some number of repetitions all of the updated syndrome terms find the $0$ value, we declare that the decoder has completed its task. We then determine if the correction recovers the encoded state to evaluate the success of the sample. In some cases, a decoder sequence will increase the weight of the syndrome. In these cases, we terminate the decoder and report a heralded failure. We discuss the runtime of our decoder and analyse the different failure mechanisms in Appendix~\ref{App:Runtime}.

\subsection{Correcting irregular lattice sizes}
\label{SubSec:Repeats}

For simplicity, we have focused on lattices of size $L \times L/2$ with $L = 2^k$ for integers $k$. However, one could generalise our method to other system sizes by duplicating a given syndrome in a regular pattern as illustrated in Fig.~\ref{Fig:Repeats}. We consider a lattice of size $L^x \times L^y $ such as that shown in the top left of Fig.~\ref{Fig:Repeats}, where we cannot find a triangular fundamental symmetry that will fit on the lattice in a regular way. To deal with this, we can duplicate the error several times, as we have shown in the top right of Fig.~\ref{Fig:Repeats}. Specifically, we duplicate the system to make a repeated lattice of size $J^x L^x \times J^y L^y$ whereby $J^x L^x =  K^x 2^{k'+1} $ and  $J^y L^y =  K^y 2^{k'} $ for some integers $J^x$, $J^y$, $K^x$, $K^y$ and $k'$ such that a periodic triangular symmetry can be obtained. Given such a symmetry, one can identify probes that identify the locations of errors in a similar way to which we have described above. Given a duplicated system that satisfies the constraints we have suggested, it may then be interesting to scale the duplicated error further to find new symmetries on a larger lattice. This may improve the error correcting power of the decoder. Similar effects have been found by studying surface codes under a specific noise model where the dimensions of the lattice are coprime~\cite{Tuckett18a}.

\section{Results}
\label{Sec:Results}

Here we present data from numerical experiments that demonstrate our decoder is robust to one-dimensional errors. We use Monte-Carlo sampling to determine the logical failure rate of the decoder for varying system sizes and error rates using two different noise models. Against an independent and identically distributed noise model, we find that the logical failure rate scales super-exponentially in $L$ for small physical error rates. The second model we test is the `spanning' noise model. This noise model introduces string-like errors along the rows and columns of the lattice. The decoder demonstrates a finite threshold for the spanning noise model. Both of these observations are consistent with the behaviour of a decoder that can correct errors with one-dimensional support.

\subsection{Independent and identically distributed noise}

\begin{figure}
	\begin{center}
		\includegraphics[width=0.5\textwidth]{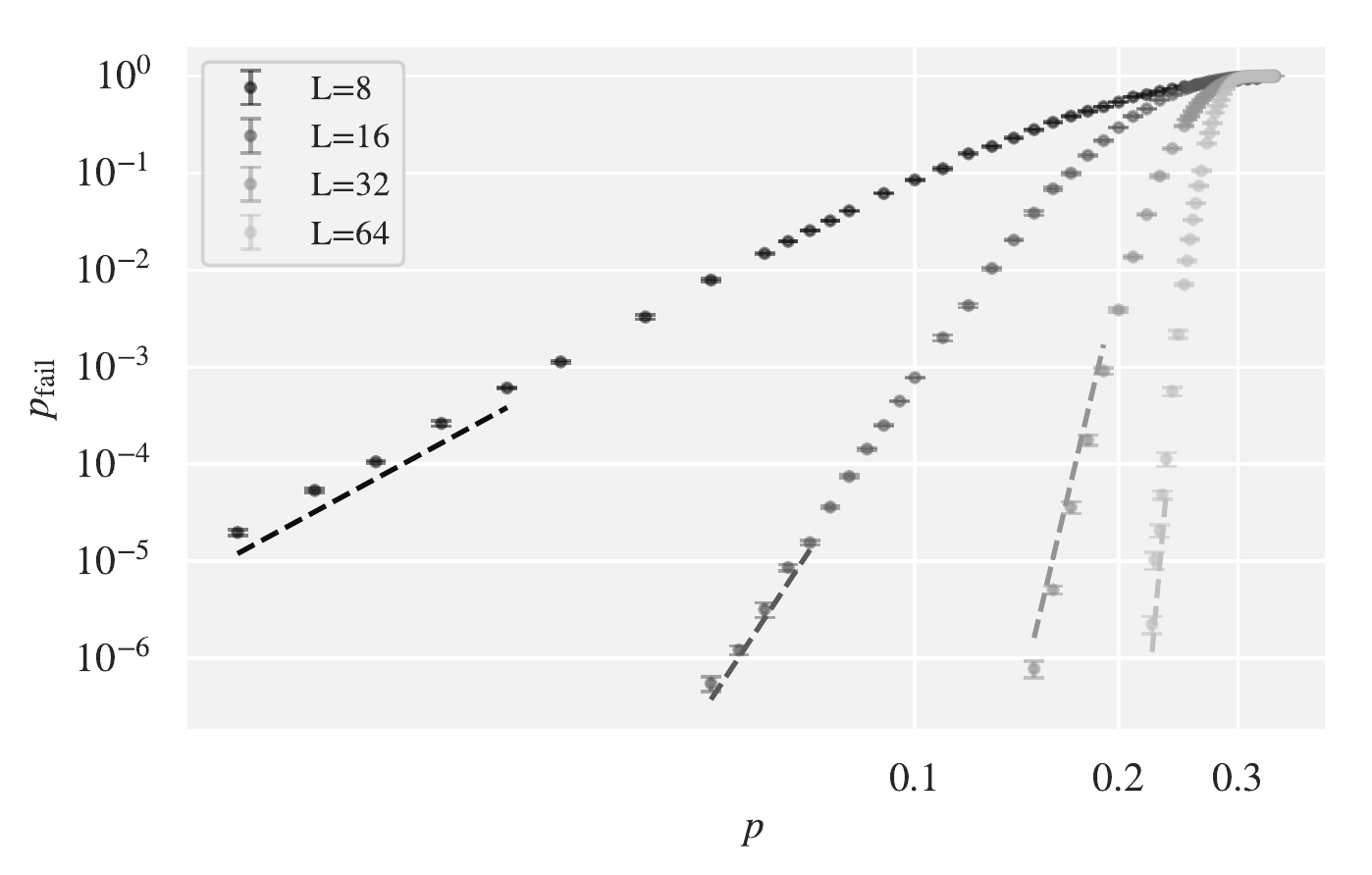}
	\end{center}
	\caption{\label{Fig:LogLogAnsatz}  Monte-Carlo sampling determines the decoder failure rate $p_{\textrm{fail}}$ for each physical error rate $p$ using an identically and independently distributed noise model. Results are shown on a logarithmic scale for system sizes $L = 8,\, 16, \,32$ and $64$ as marked in the legend. The dashed lines show the fit calculated using the ansatz of Eqn.~(\ref{Eqn:LowPhysicalErrorAnsatz}) where $A = 0.17 $, $\beta = 0.59$, $\delta = 1.25$, $\alpha = 0.18 $,  and $\gamma = 1.47$. The method used to determine these parameters is discussed in Appendix~\ref{App:Methods}. The error bars show the standard error of the mean given by the expression $\Delta p_{\textrm{fail}} = \sqrt{(1-p_{\textrm{fail}})p_{\textrm{fail}} / \eta}$ where $\eta$ is the number of Monte-Carlo samples collected. We use between $10^4$ and $10^7$ Monte-Carlo samples for each data point. For each data point shown, at least 25 failures have been obtained.
	}
\end{figure}

We begin our analysis by benchmarking our decoder using an independent and identically distributed noise model. The variable $p$ denotes the probability that a given bit is flipped. For each error we sample, we report a success if the decoder predicts the error exactly. In any other instance, the logical information has been corrupted and we report a logical failure. 

We evaluate the performance of the decoder at low error rates by comparing our data to the ansatz 
\begin{equation}
p_{\textrm{fail}} = A \exp \left(  \alpha   L^{\gamma} \log p  + \beta L^\delta \right),
\label{Eqn:LowPhysicalErrorAnsatz}
\end{equation}
where $\alpha$, $\gamma$ and $\delta$ are constants to be determined \cite{Brown16a}. Central to our analysis are $\gamma$ and $\delta$. These constants reflect the leading order scaling in $L$ of the weight of the most probable failure configurations, and the number of least weight failure configurations, respectively. Fitting to the low error data points for each system size in Fig.~\ref{Fig:LogLogAnsatz}, we find
\begin{equation}
\gamma \sim 1.47,\quad \delta \sim 1.25.
\label{Eqn:Gamma}
\end{equation}
As we will explain shortly, the values $\gamma,\,\delta >1$ are consistent with a decoder that can tolerate one-dimensional errors. We also find that $A \sim 0.17$, $\alpha \sim 0.18$ and $\beta \sim 0.59 $. The ansatz in Eqn.~(\ref{Eqn:LowPhysicalErrorAnsatz}) is superimposed with these parameters in the low error region for each system size in Fig.~\ref{Fig:LogLogAnsatz}. In Appendix~\ref{App:Methods} we explain the analysis we used to obtain these parameters.

Let us now interpret the parameters we have found using our fitting. The logical failure rate can be written explicitly with the expression $p_{\textrm{fail}} = \sum_{\sigma\in\mathcal{F}}p(\sigma)$ where $\mathcal{F}$ denotes the set of all error configurations $\sigma$ that will cause the decoder to fail, and $p(\sigma)$ is the probability that the error configuration $\sigma$ occurs with respect to the noise model. We assume that the least weight error that can cause a logical failure using our decoder has weight $t \lesssim d/2$ where $d$ is the distance of the Fibonacci code. We then write the logical failure rate as
\begin{equation}
\label{Eqn:ErrorSummation}
p_{\textrm{fail}} = (1-p)^{L^2/2} \sum_{l = t }^\infty N(l) \left(\frac{p}{  1-p}\right)^l
\end{equation} 
with respect to the independent and identically distributed noise model where $N(l)$ is the number of error configurations of weight $l$ that can cause the decoder to fail.

To determine the ability of the decoder to correct spanning errors that are introduced by the independent and identically distributed noise model, we are interested in estimating the size $t$ of the lowest weight error that could cause the decoder to fail as a function of system size. For small $p$ we assume that higher-order terms in Eqn.~(\ref{Eqn:ErrorSummation}) are negligible such that
\begin{equation}
p_{\textrm{fail}} \propto N(t) \left(\frac{p}{  1-p}\right)^t.
\end{equation} 
By taking $t \sim \alpha L^\gamma$ and $\log( N(t) ) \sim \beta L^\delta$ we recover the ansatz in Eqn.~(\ref{Eqn:LowPhysicalErrorAnsatz}) where we also suppose that $\log [p / (1-p) ] \approx \log p$ at small values of $p$.  Finding that $\gamma > 1$ is consistent with a code where the least weight error scales super linearly in $L$. Moreover, we assume that those errors may configure themselves in $\sim 2^{2t}$ different ways to produce a logical failure. Finding an entropic term $N(t)$ with $\delta > 1$ is therefore also consistent with the behaviour of a code that can tolerate one-dimensional errors.

It is also worthwhile to evaluate the threshold error rate for the independent and identically distributed noise model. The threshold error rate is the physical error rate below which the logical failure rate $p_{\textrm{fail}}$ can be suppressed arbitrarily by increasing the system size. We obtain the threshold error rate by fitting data close to the crossing
to the function 
\begin{equation}
p_{\textrm{fail}} = \exp ( B_0 + B_1 x + B_2 x^2 ), \label{Eqn:ThresholdFit}
\end{equation} 
where $x = (p - p_{\textrm{th}})L^{\mu}$, $p_{th}$ is the threshold error rate and  $B_j$ and $\mu$ are constants to be determined~\cite{Wang03}. We calculate that 
\begin{equation}
p_{\textrm{th}} \sim 0.304,
\end{equation} 
for the independent and identically distributed noise model. We also find $B_0 \sim -1.3\cdot 10^{-2}$, $B_1 \sim 2.5\cdot 10^{-3}$, $B_2 \sim -6.4\cdot 10^{-5}$ and $\mu \sim 1.7$.

Let us finally remark on the different mechanisms that cause the decoder to fail. Typically, we find that the most common failure mechanism is for the decoding sequence to terminate after an iteration increases the hamming weight of the syndrome, rather than to  predict the code state incorrectly after the successful removal of defects. The ratio between these failure mechanisms for varying system sizes is discussed in Appendix \ref{App:Runtime}.

\subsection{Spanning noise}

\begin{figure}
	\begin{center}
		\includegraphics[width=0.45\textwidth]{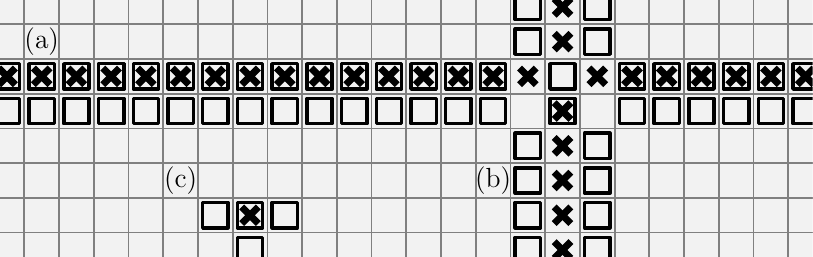}
	\end{center}
	\caption{\label{Fig:CLErrors} The spanning noise model introduces bit flips along one-dimensional horizontal~(a) and vertical~(b) lines of the lattice. We also introduce independent and identically distributed bit flips to the configuration~(c). Errors are marked by black crosses and the resulting defects are shown by hollow black squares. Bits that have been flipped twice do not contribute to the error configuration. }
\end{figure}

We assess the capability of our decoder to correct high-weight errors further using a `spanning' noise model. The noise model introduces one-dimensional errors that span the lattice. The noise model is depicted in Fig.~\ref{Fig:CLErrors}.

We introduce errors onto the codeword with bits $\sigma_{(x,y)}$ as follows.
\begin{enumerate}
\item For every $ 1 \le y \le L/2$, with probability $p$, we perform bit flips $\sigma_{(x,y)} \rightarrow  \sigma_{(x,y)} \oplus 1$  for all $ 1 \le x \le L$ with certainty, see Fig.~\ref{Fig:CLErrors}(a). \\
\item  For every $ 1 \le x \le L$, with probability $p$, we perform bit flips $\sigma_{(x,y)} \rightarrow  \sigma_{(x,y)} \oplus 1$  for all $ 1 \le y \le L/2$ with certainty, see Fig.~\ref{Fig:CLErrors}(b). \\
\item  For every face  $ f =  (x,y ) $, with probability $p$, we perform a bit flip $\sigma_{(x,y)} \rightarrow  \sigma_{(x,y)} \oplus 1$, see Fig.~\ref{Fig:CLErrors}(c).\\
\end{enumerate}

\begin{figure}
	\begin{center}
		\includegraphics[width=0.5\textwidth]{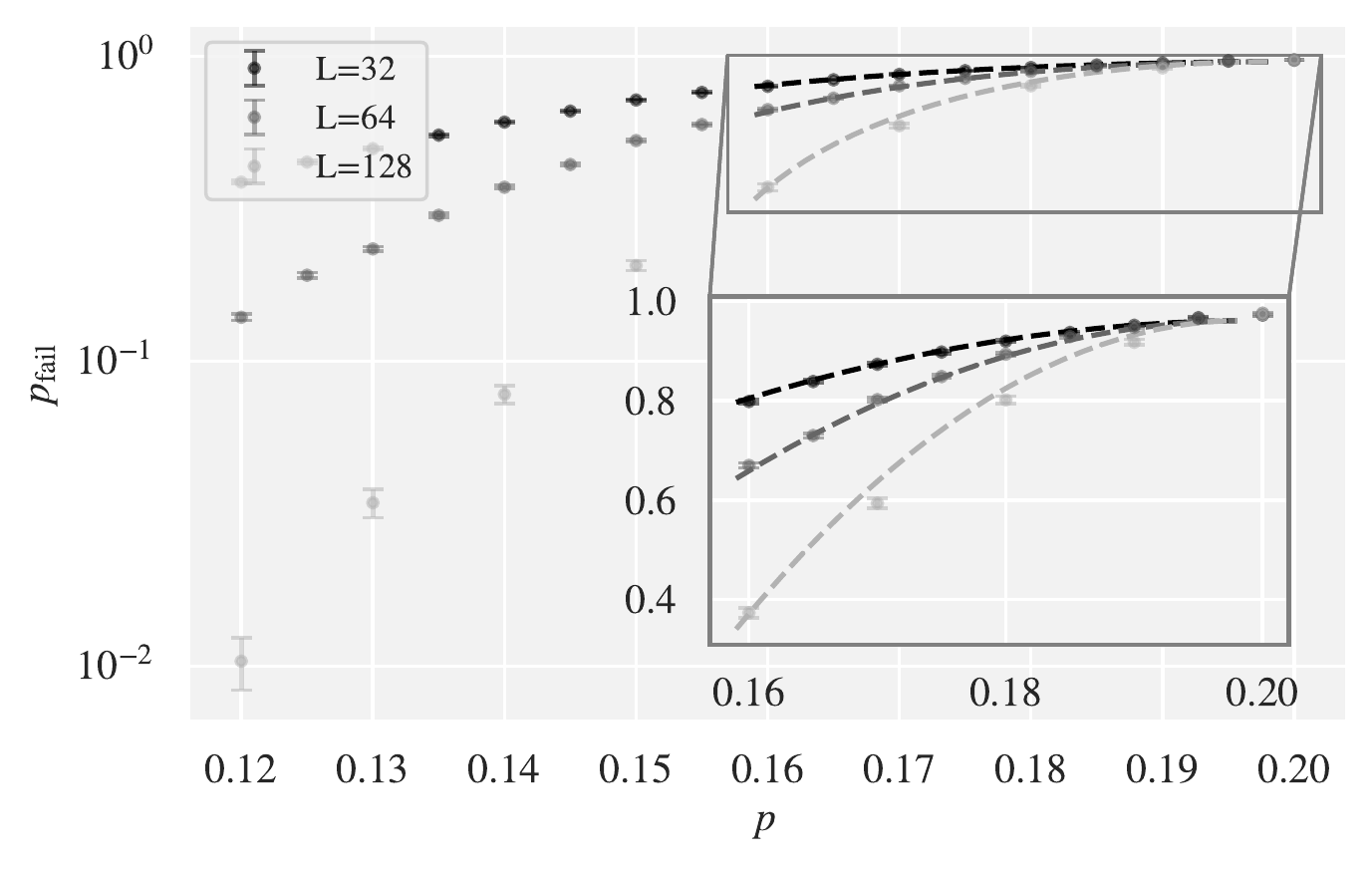}
	\end{center}
	\caption{\label{Fig:StringErrorModel} Threshold data for the spanning noise model. The logical failure rate $p_{\textrm{fail}}$ calculated with Monte-Carlo sampling is plotted as a function of the physical error rate $p$ for various system sizes $L$. The existence of a threshold at $p \sim 0.19$ is evidence that the decoder can withstand spanning errors. The dashed lines show the fit of Eqn.~\ref{Eqn:ThresholdFit} close to threshold for each system size using parameters $p_\textrm{th} = 0.197$, $B_0 =0.961$, $B_1 = 9.18\cdot 10^{-2}$, $B_2 = -3.04$ and $\mu =0.502$.}
\end{figure}

We determine the threshold error rate for the spanning noise model by fitting to Eqn.~(\ref{Eqn:ThresholdFit}). We find
\begin{equation}
p_{\textrm{sp.th.}} = 0.197,
\end{equation} 
where we fit to the data shown in Fig.~\ref{Fig:StringErrorModel} to obtain this value. Other parameters calculated are $B_0 \sim 1.0$, $B_1 \sim 9.2\cdot 10^{-2}$, $B_2 \sim -3.0$ and $\mu \sim 0.50$. The dashed lines in Fig.~\ref{Fig:StringErrorModel} show the fitting close to threshold for each system size.
	
It will be remarkable to extend this result to quantum error correcting codes. Research into quantum error correction has focused on two-dimensional codes that will invariably undergo logical failures if errors span the lattice. In future work, it will be valuable to find decoders for quantum codes~\cite{Castelnovo11, Haah11, Yoshida13a} that have the potential to tolerate spanning errors.

\section{Discussion}

We have proposed an efficient decoding algorithm for the Fibonacci code that pairs defects on fractal symmetries to determine the locations of errors. Our numerical results indicate that the code is robust to errors that span the lattice. Moving forward, it will be interesting to produce an analytical proof that a decoder is capable of decoding all spanning errors on a fractal code.

The methods we have proposed are readily adapted to other fractal codes. These codes are of fundamental interest as they have been shown to saturate the information storage bounds of local codes~\cite{Yoshida13}. Our numerics establish a lower bound on the distance of these codes. Refinements of the decoder we have presented here may help to tighten these bounds. We give a discussion on other classical codes that we find interesting in Appendix~\ref{App:OtherDecoders}.The examples we give have a variety of symmetries that we might consider exploiting to design and test new decoders using the methods we have presented here.

The leading motivation for this work was to develop methods to decode higher-dimensional fractal quantum error-correcting codes. The two-dimensional quantum error-correcting codes that we are currently developing are fundamentally vulnerable to one-dimensional errors. Should there be some process in the laboratory that introduces spanning errors to a lattice system, then it will be important to find codes that are robust to such errors. Adapting the methods we have presented here to the quantum case may provide a solution to this problem. We leave this to future work.

\appendix

\subsection{Runtime and failure mechanisms}
\label{App:Runtime}

\begin{figure}[b]
	\includegraphics[width=0.5\textwidth]{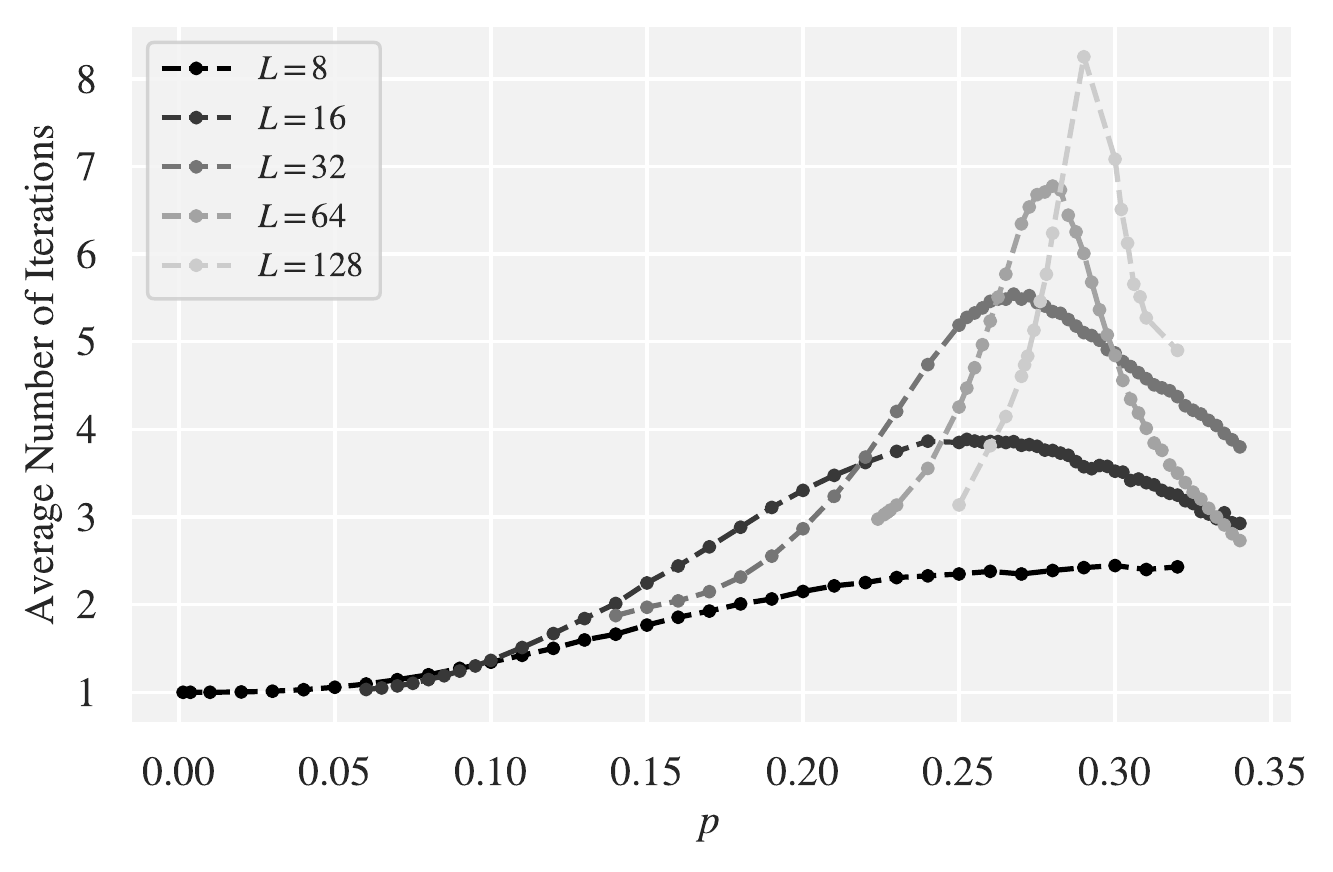}
	\caption{ \label{Fig:AverageNumberOfIterations} The average number of iterations the decoder uses for a single run.  Below $p \lesssim 0.25$, the largest system size requires the fewest iterations, indicating a large regime below threshold where the number of iterations required to complete decoding does not diverge with system size.}
\end{figure}

The runtime of the decoder is determined by the runtime of the pairing subroutines used in one iteration of the decoder, and the number of repeated iterations of these matchings that are required to neutralise the defects. We determine a correction for a single iteration by performing $L^2/2$ pairing subroutines on $L^D$ sites of the lattice that can occupy defects where $D = 1+\log_2 \phi \approx 1.69$ and $\phi$ is the golden mean~\cite{Yoshida13a, Devakul19}. By parallelising calls, the time taken to find a single iteration of the decoder is equal to the time-complexity of a single pairing subroutine. As we chose to use minimum-weight perfect matching, a single subroutine will take $V^3$ where $V = \mathcal{O} (L^D)$ is the number of graph nodes input into the subroutine. However, an iteration is readily sped up using other pairing subroutines that can find a pairing in time closer to $\mathcal{O}(L^D)$~\cite{Delfosse17}.  The Kolmogorov implementation of the minimum-weight perfect matching algorithm we use has worst case complexity of $\mathcal{O}(V^3 E)$ where $V$ is the number of input vertices and $E = \mathcal{O}(V^2)$ is the number of input edges for a complete graph.

\begin{figure}
	\includegraphics[width=0.5\textwidth]{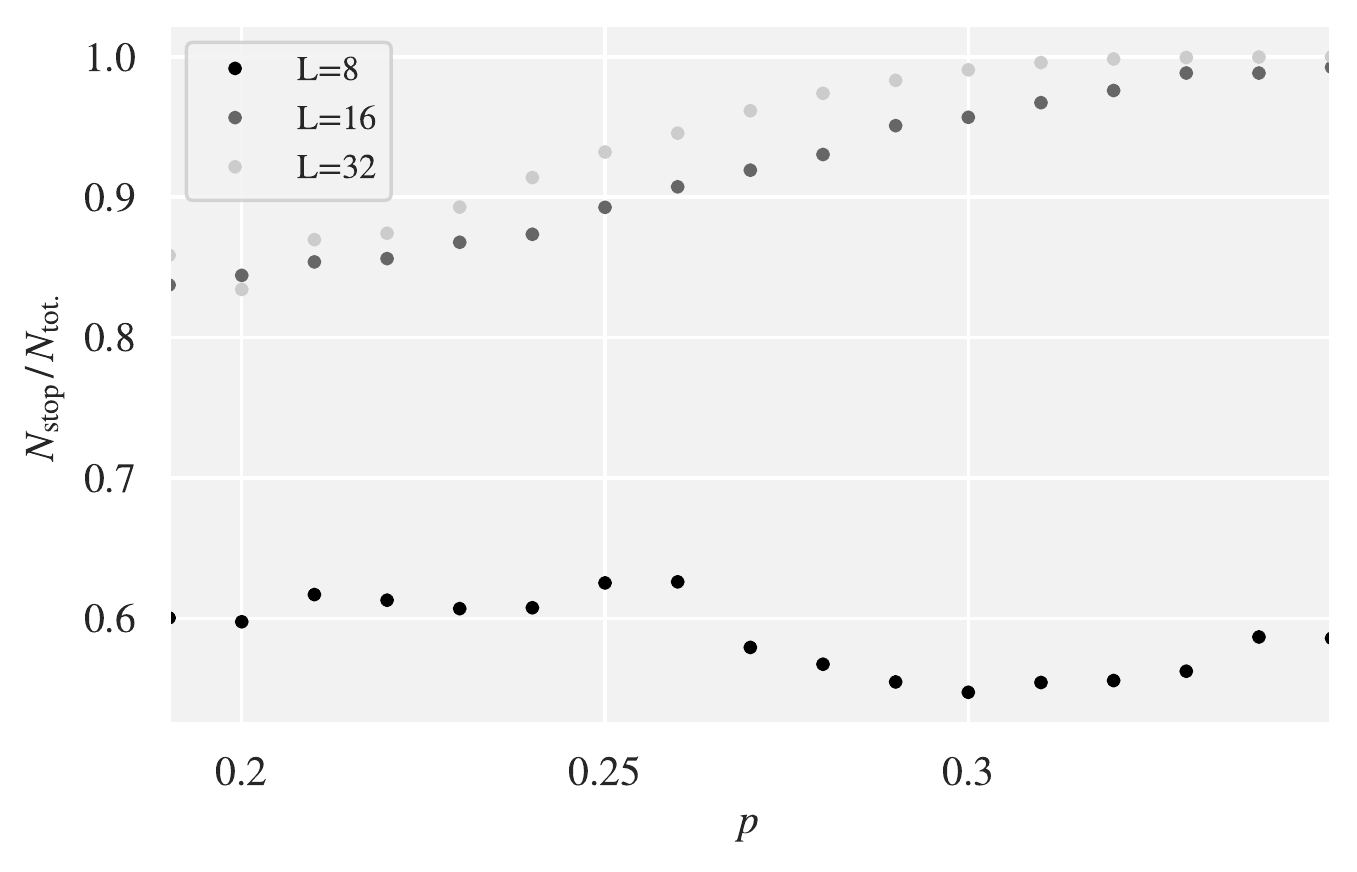}
	\caption{\label{Fig:IncorrectVsGiveUp} Plot showing the frequency of heralded failures. Our decoder can fail in two ways. Assuming a failure has occurred, it either found an incorrect codeword, or the decoder terminated due to a heralded failure. The latter occurs if one iteration of the decoder increases the hamming weight of the syndrome. We plot the ratio of the number of terminations $N_{\text{stop}}$ against the total number of failures $N_{\text{tot.}}$ as a function of error rate $p$ for different system sizes, $L$ where we obtain between $10^4 $ and $ 10^5$ failures per data point. The plot shows that as the system size increases, it becomes increasingly likely that the failure will be heralded. }
\end{figure}

It remains to determine how many times we need to reiterate the decoder before all defects are neutralised. While it is difficult to determine how well a single iteration of the decoder will perform in general, at low error rates it is reasonable to assume that more bit flips will be corrected accurately with a larger system. As such, we might expect a larger system size to require fewer iterations to achieve a correction that neutralises the syndrome.

In Fig.~\ref{Fig:AverageNumberOfIterations} we plot the average number of iterations that are required before either all defects are neutralised or the decoder terminates. We show results for different system sizes and values of $p$. We find that at $p \lesssim 0.25$ that the largest system size we study requires the fewest number of attempts on average. Our findings indicate that the number of iterations does not diverge with increasing system size for moderate physical error rates. Near to threshold, the number of iterations required by the decoder grows with system size. In practice, we would not use the code in this regime. Nevertheless, in future work it may be interesting to determine the dynamics of the decoder in this regime at larger system sizes. Here, larger system sizes are more likely to successfully correct the error by using more iterations. A better understanding of this dynamic may reveal ways to improve the decoder we have presented.

Let us finally comment on the different failure mechanisms of the decoder. The iterative decoder can produce a logical error either by incorrect code state assignment, or, a heralded failure can occur after an iteration of the decoder increases the hamming weight of the syndrome. Figure~\ref{Fig:IncorrectVsGiveUp} shows how failed Monte-Carlo samples are distributed between each of these failure mechanisms. Our results show that it is much more common for the decoder to terminate and herald a failure, than to find an incorrect codeword. In practice this feature is advantageous as we can repeat the protocol when we have knowingly failed to read the encoded information.

\subsection{Behaviour at low error rates}
\label{App:Methods}

\begin{figure}[b]
	\includegraphics[width=0.5\textwidth]{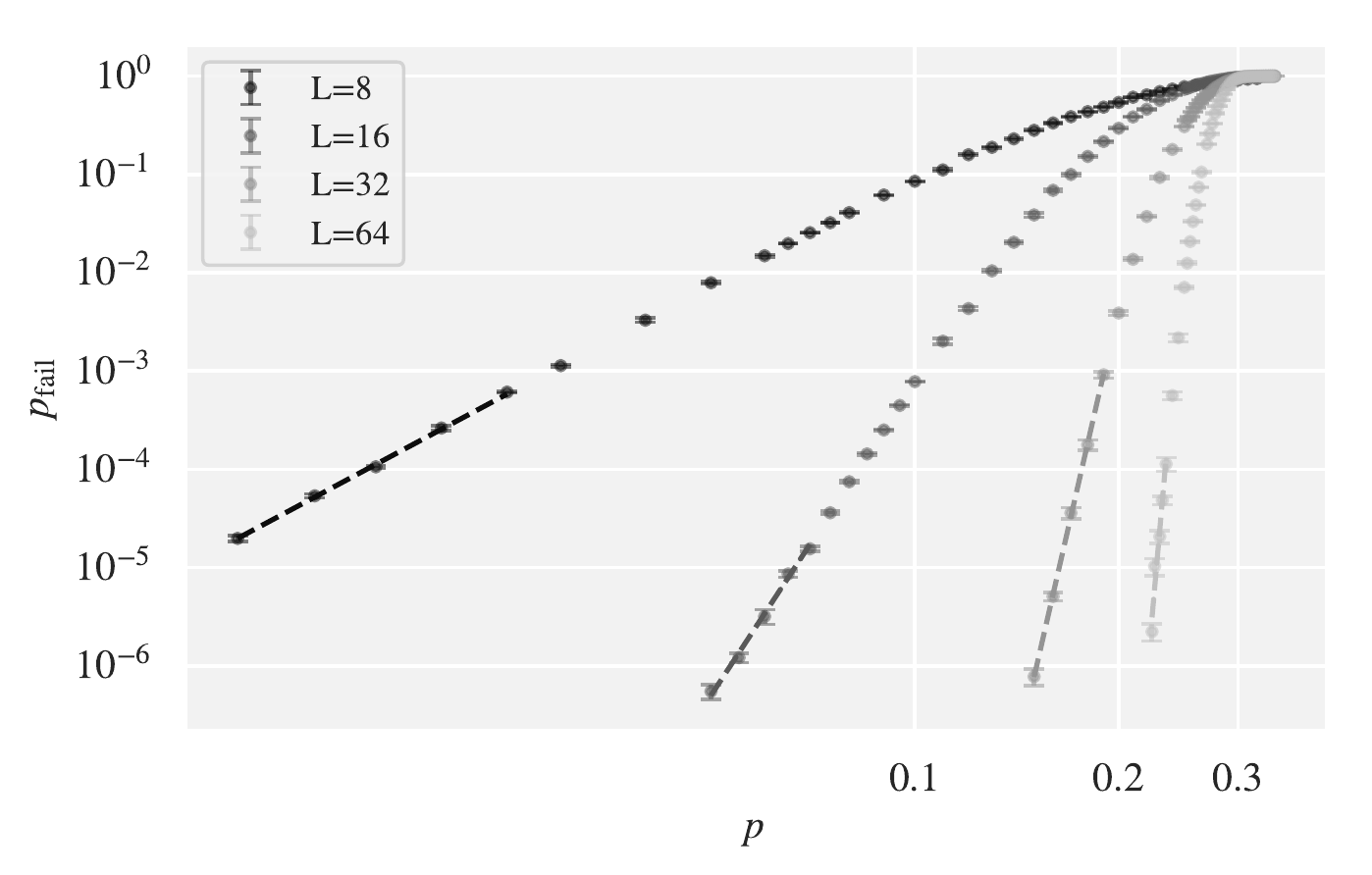}
	\caption{\label{Fig:LogLogStraightLineFit} Plot showing logical failure rate $p_{\textrm{fail}}$ as a function of physical error rate $p$ for different system sizes $L$. For each system size, we use the method of least squares to fit a straight line to the five data points with lowest $p$. These fits are shown as dashed lines in the figure. We plot gradients $G(L)$ and intercepts $I(L)$ for each system size in Figs.~\ref{Fig:GradientAgainstSystemSize} and~\ref{Fig:InterceptAgainstSystemSize}, respectively.  }
\end{figure}

\begin{figure}
	\includegraphics[width=0.5\textwidth]{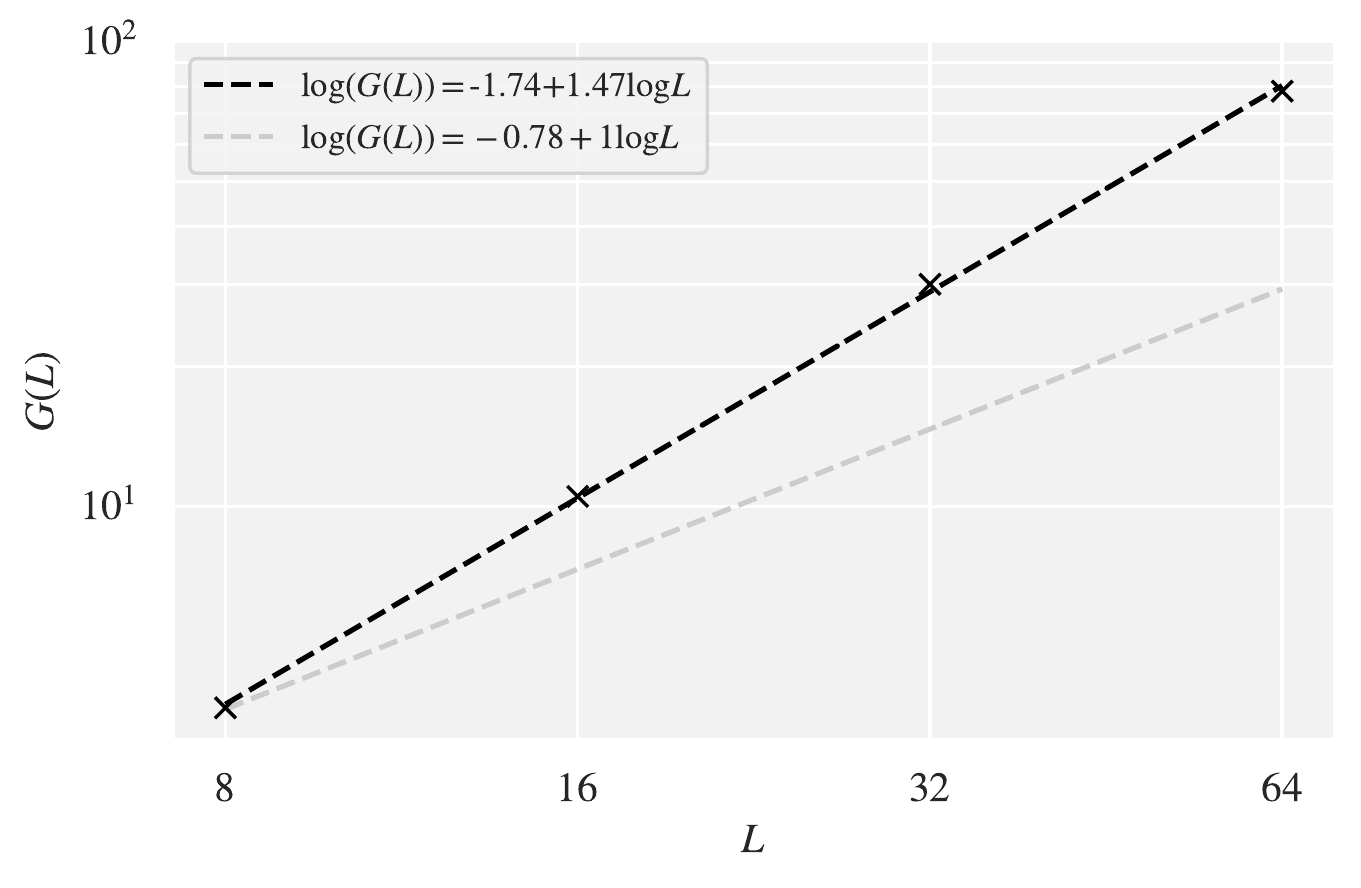}
	\caption{\label{Fig:GradientAgainstSystemSize} Gradients $G(L)$ obtained for the fittings shown in Fig.~\ref{Fig:LogLogStraightLineFit} plotted as a function of system size $L$.  We fit the data to a straight line on the logarithmic plot to obtain $\gamma$, the gradient of the fitting, and $\log \alpha$, its intercept. We use the method of least squares on our data to calculate that $\gamma = 1.47$ and $\log \alpha = -1.74$. We compare our fitting to a line with gradient $1$, as shown by the light-grey dashed line.}
\end{figure}

\begin{figure}[b]
	\includegraphics[width=0.5\textwidth]{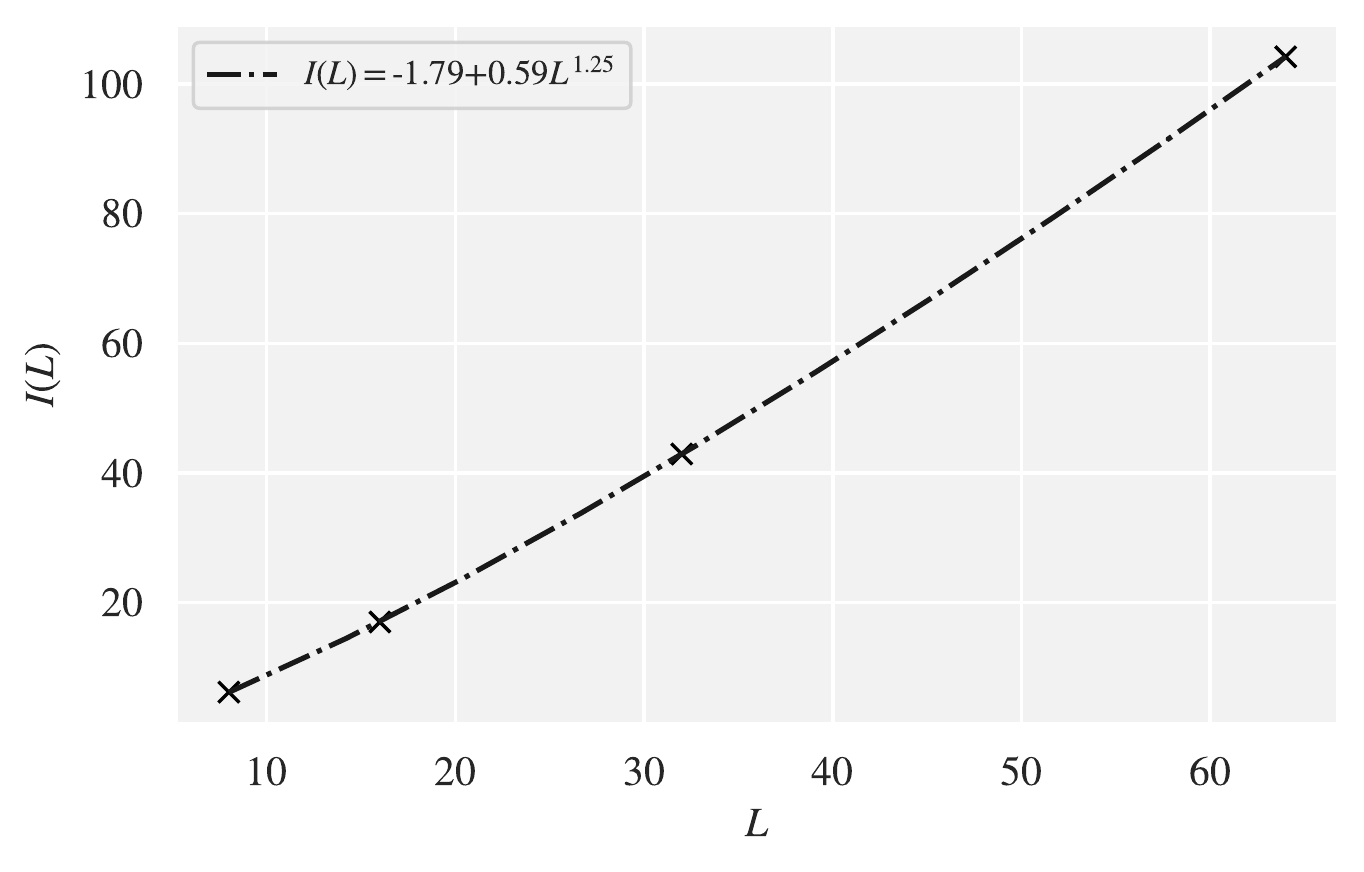}
	\caption{\label{Fig:InterceptAgainstSystemSize}  
		Plot showing intercepts $I(L)$ for the linear fittings shown in Fig.~\ref{Fig:LogLogStraightLineFit} as a function of system size, $L$. Using the Python package SciPy we find $I(L) = -1.79 + 0.59 L^{1.25}$. The fit is shown by a dashed line.
	}
\end{figure}

Here we describe how we determine the free coefficients in Eqn.~(\ref{Eqn:LowPhysicalErrorAnsatz}). We take the logarithm of both sides of Eqn.~(\ref{Eqn:LowPhysicalErrorAnsatz}) to obtain
\begin{equation}
\log p_{\textrm{fail}} = \log A + \beta L^\delta + \alpha L^{\gamma} \log p .
 \label{Eqn:LogAnsatz}
\end{equation}
For a fixed system size, we see that the gradient of a linear fit of $\log p_{\textrm{fail}}  $ plotted as a function of $\log p$ is $G(L) = \alpha L^\gamma$, and that the intercept of this line is $I(L) = \beta L^\delta + \log A$.
 We therefore find $G(L)$ and $I(L)$ for each system size separately using data points where $p$ is small. Each of the linear fittings are shown in Fig.~\ref{Fig:LogLogStraightLineFit}.
We obtain values for $\alpha$ and $\gamma$ by plotting $\log G(L)$ as a function of $\log L $, see Fig.~\ref{Fig:GradientAgainstSystemSize}. From the definition of $G(L)$ we have that
\begin{equation}
\log G(L) = \log \alpha + \gamma \log L.
\label{Eqn:LogGL}
\end{equation}
We make a linear fit to system sizes $L =8,\, 16,\,32$ and $64$ to find
\begin{equation}
\log \alpha = -1.74, \textrm{ and } \gamma = 1.47.
\end{equation}
Similarly, we obtain values of $A$, $\beta$ and $\delta$ by fitting the function $I(L)$ to our data as a function of $L$ using the Python package SciPy.
\begin{equation}
I(L) = \log A + \beta L^{\delta}.
\end{equation}
We find $\log A = -1.79$, $\beta = 0.59$, and $\delta = 1.25$, see Fig.~\ref{Fig:InterceptAgainstSystemSize}.

The data we have collected was obtained using $\sim 10^5$ CPU hours.

\subsection{Comparison with other classical codes}
\label{App:OtherDecoders}

In this Appendix we discuss some other two-dimensional classical models with higher-dimensional quantum analogues that mimic some of the encouraging features of the Fibonacci code. We first discuss the eight-vertex model; a classical model that encodes a number of logical bits that scales with the linear size of the lattice. We also discuss the two-dimensional Ising model; a model capable of correcting very high-weight errors including those that span one-dimensional bands of the lattice. The Newman-Moore model is also of interest as, like the Fibonacci model, it has similar fractal features.

\subsubsection{The eight-vertex model}

We first consider the eight-vertex model~\cite{Sutherland70, Baxter71, Baxter72, Bombin12}. One could view this model as a classical analogue of the X-cube model~\cite{Vijay16}, see Ref.~\cite{Brown19} where a decoder is introduced for the X-cube model. It is also noteworthy that this model is equivalent to the surface code that only experiences Pauli-Y errors~\cite{Tuckett18} where the star and plaquette operators of the surface code are expressed in their conventional Pauli-X and Pauli-Z bases, respectively~\cite{Kitaev03}.

\begin{figure}
\begin{center}
\includegraphics{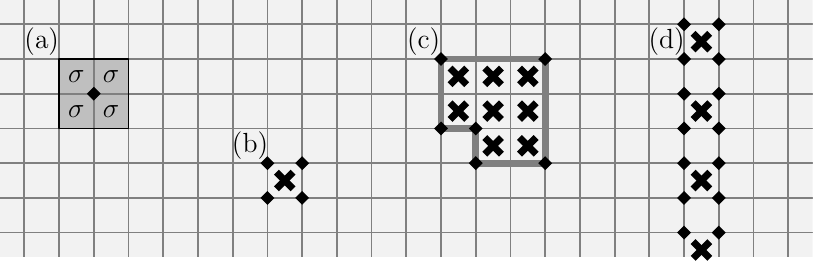}
\end{center}
\caption{ \label{Fig:EightVertex} The eight-vertex model with bits lying on the faces of a square lattice. (a)~A weight-four parity check $S_v$. Vertex $v$ is marked by a diamond in the centre of the parity check. (b)~A single error at face $f$ introduces four defects on the vertices at the corners of $f$. (c)~In general, the code introduces defects on the corners of the boundary of the error. (d)~An example of a one-dimensional error that will cause a logical failure.}
\end{figure}

The eight-vertex model can be represented on an $L\times L$ square lattice with a bit on each of its faces $f$, see Fig.~\ref{Fig:EightVertex}. The eight-vertex model has one-dimensional logical operators and has distance $L$. As such, the code is not robust to one-dimensional errors. However, like the Fibonacci code, the number of logical bits encoded by the eight-vertex model scales with $L$. It encodes $\mathcal{O} (L)$ logical bits.

The eight-vertex model has stabilizers 
\begin{equation}
S_v = \bigoplus_{\partial f \ni v} \sigma_f,
\end{equation} 
for each vertex $v$ of the lattice where $\partial f$ are the set of vertices at the corners of $f$, see Fig.~\ref{Fig:EightVertex}(a). A single error on face $f$ introduces four defects on the vertices at the corners of $f$, see Fig.~\ref{Fig:EightVertex}(b), and in general, the model introduces defects at the corners of the boundary see Fig.~\ref{Fig:EightVertex}(c). A logical error is introduced by flipping all the bits along a single vertical or horizontal line. An example of an error that cannot be corrected is shown in Fig.~\ref{Fig:EightVertex}(d).

This model can be decoded by performing minimum-weight perfect matching along the vertical and horizontal lines of the model. This is due to the one-dimensional symmetries of the model. The edges returned from the matching produce the boundary of the error. A correction is then obtained by finding the interior of the boundary. This decoder was recently implemented in Ref.~\cite{Tuckett19}.

\subsubsection{The Ising model}

We next compare the Fibonacci code with the two-dimensional Ising model~\cite{Ising25}. The Ising model is a classical analogue~\cite{Day12} of a self-correcting quantum memory~\cite{Brown16} such as the four-dimensional toric code~\cite{Dennis02, Brown16, Alicki10, Bombin13}. The model encodes a single logical bit, and has distance $L^2$. As such, like the Fibonacci code, the model should be tolerant to one-dimensional errors.

\begin{figure}[b]
\begin{center}
\includegraphics{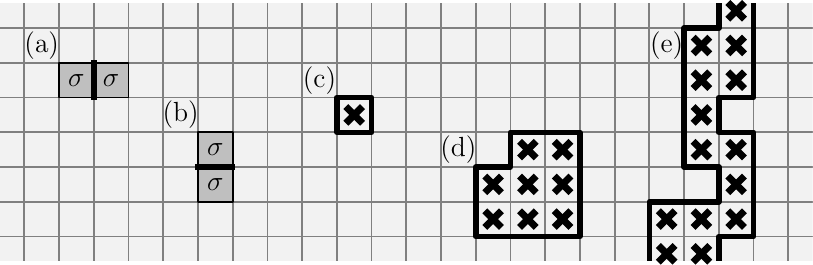}
\end{center}
\caption{\label{Fig:Ising} The two-dimensional Ising model with bits on the faces. (a)~and (b)~show the weight-two stabilizers associated to the thick black edges. (c)~A single error will introduce edge defects on the edges that bound the error. (d)~Large errors introduce defects at the edges of its boundary. (e)~Errors that span the lattice can be corrected.}
\end{figure}

We define the model with a bit on each face of an $L \times L$ square lattice. The model has a stabilizer $S_e$ associated to each edge $e$ of the square lattice, such that
\begin{equation}
S_e = \bigoplus_{\partial f \ni e} \sigma_f
\end{equation}
where $\partial f$ is the set of edges on the boundary of face $f$. We show two stabilizers in Figs.~\ref{Fig:Ising}(a) and~(b). Defects lie on the edges of the model. A single error introduces defects on the four edges that bound the error, see Fig.~\ref{Fig:Ising}(c). In general, a large error introduces defects on all the edges of its boundary, see Fig.~\ref{Fig:Ising}(d). The code can correct one-dimensional errors such as that shown in Fig.~\ref{Fig:Ising}(e).

The Ising model can be decoded using majority vote. The faces of the lattice are divided into two subsets that are separated by the edge defects that describe the boundary of the error. We correct the lattice by flipping all bits on the smallest of the two subsets of faces. This global decoder will have a very high threshold.

The two-dimensional Ising model also has local symmetries and can therefore be decoded with cellular automata ~\cite{Grinstein04}, even if some of the stabilizer measurements are unreliable. However, a cellular automaton decoder will be unable to decode errors that span the lattice. Cellular automata automaton decoders for quantum self-correcting memories are studied in Refs.~\cite{Dennis02, Pastawski11, Breuckmann17, Kubica19a}.

Furthermore, as the model is self-correcting, it is also capable of single-shot error correction~\cite{Bombin15a} where we suppose that stabilizer measurements are unreliable. This means that the Ising model is robust to both one-dimensional errors and time-correlated errors~\cite{Bombin16a}. As such, the code is still functional, even if some stabilizers malfunction permanently over the lifetime of the code.

\subsubsection{The Newman-Moore model}

\begin{figure}
\begin{center}
\includegraphics{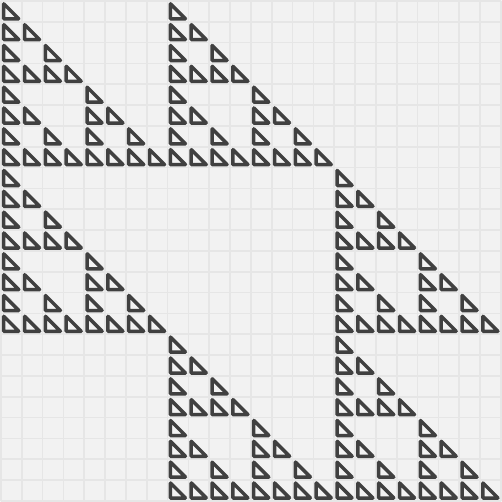}
\end{center}
\caption{A square lattice of linear size $L = 3\cdot 2^k$ with $k = 3$. A triangle is drawn for each face $f$ if $S^{\textrm{NM}}_f$ is a member of the symmetry $\Sigma$, as described in the main text. \label{Fig:Sierpinski}}
\end{figure}

The Newman-Moore model~\cite{Newman99} and its higher-dimensional generalisations~\cite{Yoshida13} are the canonical example of classical fractal codes. In two dimensions, it has stabilizers of the form
\begin{equation}
S_f^{\textrm{NM}} = \sigma_{f} \oplus \sigma_{f+\hat{x}} \oplus \sigma_{f+\hat{y}}.
\end{equation}
To delocalise three defects over the two-dimensional variant of this model, bit flips are introduced onto a collection of lattice sites with the support of a Sierpinski triangle. As such, the physics of these systems is similar to that of the Fibonacci code that has been the focus of this work.

It may be of value to design a decoder for this code, as this exercise may reveal new insights into decoding quantum versions of fractal codes~\cite{Haah11}. Moreover, comparison of this code with the Fibonacci code studied here may give us new insights into the role that entropy plays in the performance of error correction~\cite{Beverland18}.

One could conceive of using the methods we have presented to design a decoder for this code. We show a symmetry of the two-dimensional code in Fig.~\ref{Fig:Sierpinski}, where we draw a triangle for each face $f$ whereby $S_f^{\textrm{NM}} \in \Sigma$ for some symmetry, $\Sigma$. The symmetry can be divided into six distinct Sierpinski triangles. One can see from the figure that there are nine distinct locations where a single error will change the parity of defects lying on the stabilizers of the separate triangles. Using these, one can come up with three unique tests to determine if physical bits have experienced errors.

\section*{Acknowledgment}
	We are grateful to N. Breuckmann, S. Bartlett, C. Dawson, L. Donini, S. Flammia, M. Kesselring, A. Kubica, F. Pastawski, J. Preskill, A. Robertson, K. Sahay, U. Schneider, W. Shirley and D. Tuckett for helpful and encouraging discussions. We are especially thankful to D. Williamson for introducing us to the Fibonacci code, and to N. Delfosse and N. Nickerson for comments on an earlier version of our draft. GMN is grateful for the hospitality of the School of Physics at the University of Sydney. The authors acknowledge the facilities of the Sydney Informatics Hub at the University of Sydney and, in particular, access to the high performance computing facility Artemis. This work is supported by the Australian Research Council via the Centre of Excellence in Engineered Quantum Systems (EQUS) project number CE170100009. GMN is currently supported by the Cambridge Trust and the Royal Society Te Ap\={a}rangi - Rutherford Foundation.  BJB is also supported by the University of Sydney Fellowship Programme.

 \bibliographystyle{unsrt}
 \bibliography{QECreferences}

\begin{IEEEbiography}[{\includegraphics[width=1in,height=1.25in,clip,keepaspectratio]{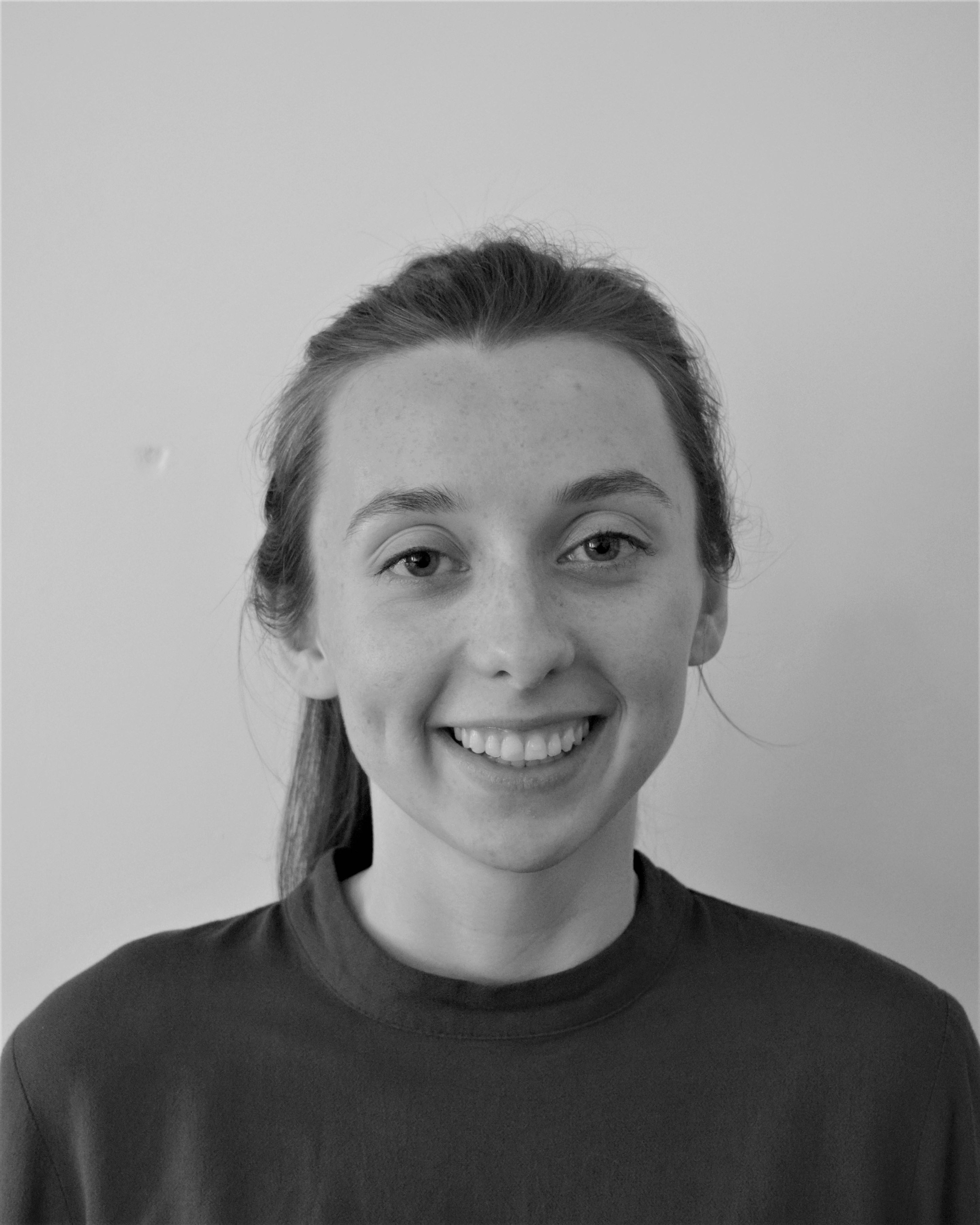}}]{Georgia Nixon}
Georgia M. Nixon was born in Paddington, Sydney, Australia. She  received a B.Sc. degree in physics and mathematics and a B.Mus. in classical performance piano from the University of Otago, New Zealand, in 2017 and a B.Sc. (Honours) degree in physics from the University of Auckland, New Zealand, in 2018. In 2019, she was a Research Assistant at the School of Physics, University of Sydney, Australia as a member of the Centre for Excellence in Engineered Quantum Systems (EQUS). She is currently pursuing a Ph.D. at the Cavendish Laboratory, University of Cambridge, UK. Her research interests include many-body quantum systems, topological phases of matter and quantum information. 
Ms. Nixon was the recipient of the Cambridge Rutherford Memorial Ph.D. Scholarship in 2019.
\end{IEEEbiography}

\begin{IEEEbiography}[{\includegraphics[width=1in,height=1.25in,clip,keepaspectratio]{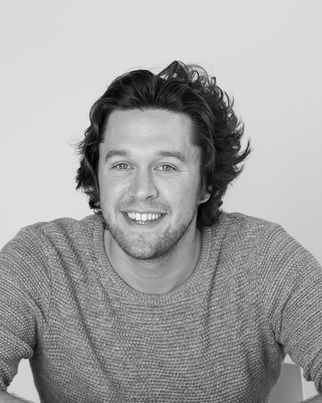}}]{Benjamin Brown}
Benjamin J. Brown was born in King's Lynn, Norfolk, UK, in 1988. He received a B.Sc. with honours in mathematics and physics from the University of Leeds, UK, in 2009, an M.Res. in controlled quantum dynamics in 2010 from Imperial College London, UK, and a Ph.D. in quantum information in 2014, also from Imperial College London.

He is currently working as a Senior Research Associate in the School of Physics at the University of Sydney, Australia as a member of the Centre for Excellence in Engineered Quantum Systems(EQUS) supported by the Australian Research Council. Previously he held positions as a Postdoctoral Researcher at the Niels Bohr Institute at the University of Copenhagen, Denmark and as a Research Associate in the Department of Physics at Imperial College London, United Kingdom. His research interests include quantum information, quantum error correction, fault-tolerant quantum computing and topological phases of matter.

Dr. Brown's research awards include the Marie Sk\l odowska Curie Award (European Commission), the University of Sydney Fellowship (University of Sydney, Deputy Vice Chancellor Research), and the Doctoral Prize Fellowship (Engineering and Physical Sciences Research Council, United Kingdom).
\end{IEEEbiography}

\end{document}